\documentclass[fleqn,11pt]{article}
\usepackage{amscd,amsmath,amssymb,verbatim}
\usepackage{amsthm}
\usepackage[pdftex]{graphicx}
\usepackage{epstopdf}
\usepackage[T1]{fontenc}
\usepackage{ae,aecompl}
\usepackage{subcaption}
\usepackage{array}
\usepackage{mathrsfs}
\usepackage[nohead]{geometry}
\usepackage[singlespacing]{setspace}
\usepackage{rotating}
\usepackage{multirow}
\usepackage{threeparttable}
\usepackage{enumerate}
\usepackage{placeins}
\usepackage{color, colortbl}
\definecolor{blue}{rgb}{0,0,0}
\usepackage{accents}
\usepackage{pdflscape}
\usepackage{longtable}
\usepackage{natbib}
\usepackage{graphicx}
\usepackage{subcaption}
\usepackage[T1]{fontenc}
\usepackage{lmodern}
\usepackage{textcomp} 

\usepackage{float}
\numberwithin{equation}{section}

\newcommand{\xvec}{\boldsymbol}
\newcommand{\xmat}{\mathbf}

\usepackage{hyperref}

\makeatletter
\newcommand{\distas}[1]{\mathbin{\overset{#1}{\kern\z@\sim}}}%
\newsavebox{\mybox}\newsavebox{\mysim}
\newcommand{\distras}[1]{%
  \savebox{\mybox}{\hbox{\kern3pt$\scriptstyle#1$\kern3pt}}%
  \savebox{\mysim}{\hbox{$\sim$}}%
  \mathbin{\overset{#1}{\kern\z@\resizebox{\wd\mybox}{\ht\mysim}{$\sim$}}}%
}
\makeatother
\newtheorem{thm}{Theorem}[section]

\renewcommand{\hat}[1]{\widehat{\text{$#1$}}}
\makeatletter
\newsavebox\myboxA
\newsavebox\myboxB
\newlength\mylenA

\renewcommand*\bar[2][0.85]{%
    \sbox{\myboxA}{$\m@th#2$}%
    \setbox\myboxB\null
    \ht\myboxB=\ht\myboxA%
    \dp\myboxB=\dp\myboxA%
    \wd\myboxB=#1\wd\myboxA
    \sbox\myboxB{$\m@th\overline{\copy\myboxB}$}
    \setlength\mylenA{\the\wd\myboxA}
    \addtolength\mylenA{-\the\wd\myboxB}%
    \ifdim\wd\myboxB<\wd\myboxA%
       \rlap{\hskip 0.5\mylenA\usebox\myboxB}{\usebox\myboxA}%
    \else
        \hskip -0.5\mylenA\rlap{\usebox\myboxA}{\hskip 0.5\mylenA\usebox\myboxB}%
    \fi}
\makeatother
\makeatletter
\def\@biblabel#1{\hspace*{-\labelsep}}
\makeatother
\hoffset 
\voffset
\author{Ariane N. Meli Chrisko\thanks{University of Glasgow, UK, email: a.meli-chrisko.1@research.gla.ac.uk} \and Philipp Otto\thanks{University of Glasgow, UK, email: philipp.otto@glasgow.ac.uk, Corresponding author}  \and Wolfgang Schmid\thanks{European University Viadrina, Frankfurt (Oder), Germany, email: schmid@europa-uni.de}  }\medskip

\date{\today}
\title{Exponential Spatiotemporal GARCH Model with Asymmetric Volatility Spillovers}

\begin{document}
\maketitle
\sloppy

\singlespacing

\begin{abstract}
\noindent 
This paper introduces a spatiotemporal exponential generalised autoregressive conditional heteroscedasticity (spatiotemporal E-GARCH) model, extending traditional spatiotemporal GARCH models by incorporating asymmetric volatility spillovers, while also generalising the time-series E-GARCH model to a spatiotemporal setting with instantaneous, potentially asymmetric volatility spillovers across space. The model allows for both temporal and spatial dependencies in volatility dynamics, capturing how financial shocks propagate across time, space, and network structures. We establish the theoretical properties of the model, deriving stationarity conditions and moment existence results. For estimation, we propose a quasi-maximum likelihood (QML) estimator and assess their finite-sample performance through Monte Carlo simulations. Empirically, we apply the model to financial networks, specifically analysing volatility spillovers in stock markets. We compare different network structures and analyse asymmetric effects in instantaneous volatility interactions.\\
Word count: 6,469

\end{abstract}

\onehalfspacing



\section{Introduction}

Volatility modelling is crucial for understanding and managing financial market risk. Classical time-series volatility models like GARCH models capture the clustering of volatility over time and GARCH models have a long history in financial research \citep{Engle82,bollerslev1986generalized,andersen1998answering, francq2019garch}, but modern financial markets are highly interconnected across assets and regions. Typically, the dimensionality of financial markets is too high to be effectively managed by multivariate time-series GARCH models that incorporate instantaneous cross-sectional dependence by the (full) covariance matrix of the errors, such as multivariate GARCH models, constant conditional correlation (CCC) or dynamic conditional correlation (DCC) models \citep{bauwens2006multivariate, bollerslev1990modelling, Engle02}. Moreover, we typically observe that the risk of an asset contemporaneously depends on the risks of adjacent assets in the network (i.e., similar or ``nearby'' assets) in an autoregressive GARCH-like manner \citep{mattera2024network}. Thereby, the volatility of ``nearby'' stocks is more similar than for stocks further apart (see also First Law of Geography, \citealt{Tobler70}). Extending heteroscedasticity models to the spatiotemporal domain, \cite{otto2022general} introduced a unified framework that integrates both temporal and spatial volatility clustering. 

At the same time, volatility is well known to respond asymmetrically to news: negative shocks (bad news) typically increase future volatility more than positive shocks of equal magnitude (good news), a phenomenon known as the leverage effect \citep{bekaert2000asymmetric,bensaida2019good}. Traditional GARCH models do not inherently capture this asymmetry, prompting the development of models like the exponential GARCH (E-GARCH) by  \cite{Nelson91} specifically to allow positive and negative shocks to have differential impacts on volatility. These E-GARCH models also have been extended to continuous-time E-GARCH models by \cite{haug2007exponential}. Given the interconnected nature of financial networks and the presence of asymmetric responses to shocks, there is a strong motivation to develop models that jointly capture spatiotemporal volatility spillovers and asymmetric effects.

Research on volatility spillovers and spatial volatility modelling has grown in recent years (see \citealt{otto2024spatial} for a review). This paper introduces a novel exponential spatiotemporal GARCH (E-GARCH) framework that bridges two existing methodologies: the time series E-GARCH model \citep{Nelson91}, known for capturing asymmetric volatility spillovers, and spatiotemporal GARCH models \citep{otto2022dynamic,otto2018generalised}, which account for contemporaneous spatial and temporal dependencies in the volatility. Many real-world economic systems exhibit spatial or spatiotemporal dependence, as economic data is often observed at multiple geo-referenced locations or across networks, where the volatility \textcolor{blue}{at a} location may be influenced by that of its neighbours. To model these dependencies, spatial and spatiotemporal volatility models have recently been developed. Although first mentioned only as a by-product in \citet{Bera04}, formal spatial ARCH models were introduced by a \citet{otto2018generalised, otto2019stochastic}. Independently, \citet{sato2017spatial} proposed spatial log-ARCH models, designed to capture ARCH-like dependencies in the log-volatility of spatial processes. These models laid the foundation for a broader class of spatial and spatiotemporal volatility models \citep[see]{otto2024spatial}.

The remainder of the paper is organised as follows. Section \ref{sec:model} introduces the specification of the exponential spatiotemporal GARCH model, and we analyse key stochastic properties of the process in Section \ref{sec:egarch_properties}. Further, we discuss estimation and inference based on the log-likelihood function in Section \ref{sec:egarch_inference}. In Section \ref{sec:MC}, we report the results of several numerical examples and a Monte-Carlo simulation on the performance of the estimator for finite samples. The methods are eventually applied on three financial network examples from two European and one US stock market. We compare different network weight matrices and interpret the estimated asymmetric spillover effects from an economic/financial perspective. Section \ref{sec:conclusion} concludes the paper, summarising the findings and suggesting directions for future research in spatiotemporal volatility modelling with asymmetry.

\section{Model Specification}\label{sec:model}

The proposed spatiotemporal E-GARCH model can be applied in spatiotemporal settings as well as for random processes observed on (static) networks. In our setting, we consider a discrete and finite set of locations or vertices/nodes in a network, denoted as $\xvec{s}_1, \ldots, \xvec{s}_n$, and define the vector of observations at time $t$ and at all locations/on all nodes as $\xvec{Y}_t = \left(Y_t\left(\xvec{s}_i\right)\right)_{i = 1, \ldots, n}$. Throughout the remainder of the paper, we will use locations and nodes interchangeably. The observed process is defined as
\begin{equation}\label{eq:initial}
\xvec{Y}_t = \text{diag}(\xvec{h}_t)^{1/2} \xvec{\varepsilon}_t ,
\end{equation}
where $\xvec{h}_t = (h_t(\xvec{s}_1), \ldots, h_t(\xvec{s}_n))'$ represents \textcolor{blue}{a measure of the} volatility that incorporates both time and space interactions, and $\xvec{\varepsilon}_t = (\varepsilon_t(\xvec{s}_1), \ldots, \varepsilon_t(\xvec{s}_n))'$ is an independent and identically distributed error with mean zero and unit variance. To define the vector $\xvec{h}_t$ of local volatilities at time $t$, \cite{otto2022general} introduced a known function $f$ that relates $\xvec{h}_t$ to a vector $\xvec{F}_t = (f(h_t(\xvec{s}_1)), \ldots, f(h_t(\xvec{s}_n)))^\prime$, which depends on (a) nearby or adjacent observations of $f(h_t(\cdot))$, $g(Y_t(\cdot))$ or $g(\varepsilon_t(\cdot))$ for selected transformations $g$ (spatial ARCH/GARCH effects), (b) past observations $f(h_{t-1}(\xvec{s}))$, $g(Y_{t-1}(\xvec{s}))$ or $g(\varepsilon_{t-1}(\xvec{s}))$ at the same locations $\xvec{s}$ (own past, temporal ARCH effects), and (c) nearby or adjacent past observations $f(h_{t-1}(\cdot))$, $g(Y_{t-1}(\cdot))$ or $g(\varepsilon_{t-1}(\cdot))$ (spatiotemporal spillover or diffusion effect).  This general approach has the advantage that different, more complex link functions could be used or $f$ could potentially be estimated in a non-parametric way. Common choices are the identity link (spatial, spatiotemporal ARCH/GARCH models, see \citealt{otto2018generalised,otto2022general}, among others) or a logarithmic link (spatial, spatiotemporal log-ARCH/GARCH models, see \citealt{sato2017spatial,otto2022dynamic,dougan2023bayesian}, among others).

For the spatiotemporal E-GARCH model (spE-GARCH), we choose a logarithmic link function, i.e., $f(x) = \ln(x)$. In the following let $\ln(\xvec{h}_t)$ be the vector of all element-wise logarithms of $\xvec{h}_t$, i.e. $\ln(\xvec{h}_t) = \left( \ln( h_t(\xvec{s}_i) \right)_{i=1,\cdots, n}$. Then, the local (log-)volatility process is given by : 
\begin{eqnarray}\label{eq:unified2}
	\ln( \xvec{h}_t)  & = &   \xvec{\alpha_1} + \rho_0 \xmat{W}_1 {g}(\xvec{\varepsilon}_t) +  \rho_1 {g}(\xvec{\varepsilon}_{t-1}) + \lambda_0\xmat{W}_2 \ln( \xvec{h}_t)   + \lambda_1 \ln( \xvec{h}_{t-1}).
 \end{eqnarray}
Unlike the standard temporal E-GARCH model, where the conditional volatility depends solely on past shocks $\xvec{\varepsilon}_v$ for $v < t$, our spatiotemporal E-GARCH formulation also incorporates contemporaneous spatial effects through $\xvec{\varepsilon}_t$. Specifically, the term $\rho_0 \xmat{W}_1 g(\xvec{\varepsilon}_t)$ introduces an instantaneous spatial dependence, meaning that volatility at location $\xvec{s}_i$ is influenced not only by past innovations but also by the current-period shocks of neighbouring locations or adjacent nodes in a financial network. While this enhances the model’s ability to capture real-time volatility spillovers across space/network, it also introduces additional complexity in the theoretical analysis.

The parameter vector $\xvec{\alpha_1} \in \mathbb{R}^n$ defines the time-invariant node-specific effect. The real parameters $\rho_0$, $\lambda_0$ capture contemporaneous spatial effects whereas $\rho_1$ and $\lambda_1$  measure the temporal effects. The structure of the spatial dependence is defined by the $n \times n$ spatial weight matrices $\xmat{W}_{1}$ and $\xmat{W}_{2}$ for the asymmetric variance spillover effects and the spatial GARCH effects, respectively. These matrices are assumed to be known and non-stochastic. In a network setting, these matrices are adjacency matrices of edge weights, without self-loops (i.e., diagonal entries are supposed to be zero). In contrast to multivariate GARCH models, such as BEKK-GARCH or CCC and DCC models, this spatiotemporal model allows for instantaneous spatial spillovers from neighbouring locations (without any temporal delay) in a GARCH-like sense. 

The typical asymmetric spillovers of the spatiotemporal E-GARCH model are due to the definition of the function $g$ and $g(\xvec{\varepsilon}_t) = \left( g(\varepsilon_t(\xvec{s}_i) \right)_{i=1,..,n}$. Specifically, 
\begin{equation}\label{defg}
	g(\xvec{\varepsilon}_{t}) = \Theta \xvec{\varepsilon}_{t} + \xi (|\xvec{\varepsilon}_{t}| - E(|\xvec{\varepsilon}_{t}|))
\end{equation}
for all $i = 1, \ldots, n$, $|\xvec{\varepsilon}_{t}|$ denotes the vector of all absolute entries of $\xvec{\varepsilon}_{t}$, $|\Theta|<\xi$ \footnote{\textcolor{blue}{This restriction is related to the continuous invertibility condition for EGARCH processes, which ensures the stability and well-definedness of the volatility recursion; see \citet{wintenberger2013continuous} for details.}} where $\Theta$ account for the leverage effect and $\xi$ is considered as the ARCH term. Choosing $g$ as in (\ref{defg}) the model parameters $(\rho_0, \rho_1, \Theta, \xi)$ are not uniquely identified since it holds for all $x \neq 0$ and with $g(\xvec{\varepsilon}_t) = g_t(\Theta, \xi)$ that
\[ \rho_0 x \xmat{W}_1 g_t(\Theta/x, \xi/x) + \rho_1 x g_{t-1}(\Theta/x, \xi/x) = \rho_0 \xmat{W}_1 g_t(\Theta, \xi) + \rho_1 g_{t-1}(\Theta, \xi) . 
\] 
To overcome this problem, we put $\xi =1$, which is typically done for E-GARCH models.

In general, if we assume $\rho > 0$ and $\Theta < 0$, and further consider a negative shock $\varepsilon_t < 0$, then the log-volatility $\ln(\xvec{h}_t)$ becomes negative, as $\Theta - \xi < 0$. This indicates that negative shocks lead to larger increases in volatility compared to positive shocks of the same magnitude. Specifically, in a spatial and temporal context, the conditions $\rho_0\Theta < 0$ (for contemporaneous spatial effects) and $\rho_1\Theta < 0$ (for temporal propagation) imply that negative shocks induce stronger volatility spillovers both across locations and over time. This means that if a negative shock occurs at a given location, it not only increases the volatility at that location but also spreads to neighbouring locations (via $\rho_0$) and/or persists over time (via $\rho_1$), amplifying risk in a network sense. More formally, since $\Theta < 0$, multiplying by positive $\rho_0$ and $\rho_1$ ensures that negative shocks reduce $\ln(\xvec{h}_t)$ more than positive shocks would, leading to disproportionate volatility amplification. Moreover, $g$ could be specified in different ways for the spatial and temporal effects, say $g_1$ and $g_2$ with different degrees of asymmetry, which could be the subject of future research. \textcolor{blue}{For this paper, we assume that the degree of asymmetry is identical across space and time. The specification can be extended in many ways, for example, by including additional temporal lags, as in:
\begin{equation}
\log h_{t,s_i} = \alpha_1 
+  \rho_0 \sum_{j=1}^{n} w_{ij} g(\varepsilon_{t, s_j}) 
+  \sum_{k=1}^{q} \rho_k g(\varepsilon_{t-k, s_i}) +\lambda_0 \sum_{j=1}^{n} w_{ij} \log h_{t,s_j}+ \sum_{m=1}^{p}\lambda_i \log h_{t-m,s_i}
\end{equation}
with $w_{ij}$, the weights from known matrices reflecting the spatial dependence and the asymmetry term $g(\varepsilon_{t, s}) = \theta \varepsilon_{t, s} + \xi (|\varepsilon_{t,s}| - \mathbb{E}(|\varepsilon_{t,s}|))$, and \(p\) is the lag order for the volatility term and \(q\) is the lag order for the shock term \(g(\cdot)\). Moreover, different weight structures could be considered for each term, i.e., each temporal lag and the instantaneous interactions.}

The asymmetry we aim to capture in this model stems from the leverage effect, where investors react more strongly to negative news than to positive news of equal magnitude. This effect, widely observed in financial markets, arises because negative price movements increase perceived risk, prompting higher required returns, tighter margin constraints, and portfolio adjustments that further amplify volatility (see, e.g., \citealt{Day92,Engle93,Heynen94}). By incorporating spatial and temporal asymmetries, our model provides a framework to better understand how localised financial shocks propagate through financial networks\footnote{Moreover, it is worth noting that the models are scalable to larger networks with many nodes, because of the assumed GARCH-like network interaction structure, compared to multivariate GARCH models or CCC and DCC models, where the instantaneous interactions would be encoded in the full covariance matrix of the errors. This structural assumption leads to a better scalability for large networks, while the GARCH-type structure appears to be a reasonable choice in many situations.}. 

Spatial asymmetry in volatility can arise when economic and financial conditions vary across regions or when certain nodes in a financial network exhibit heightened risk exposure. For example, if we consider the returns of stocks in different regions, the volatility of returns may depend on the economic conditions of the region \citep[see, e.g.,][]{corradi2013macroeconomic,miled2022spatial}. In regions with weaker economic conditions, stock return volatility tends to be higher due to greater uncertainty, reduced market liquidity, and heightened sensitivity to external shocks. Similarly, in a network setting, certain highly interconnected or systemically important nodes may experience greater risk fluctuations, which can asymmetrically influence volatility across the network \citep{billio2012econometric,barunik2016asymmetric}. Moreover, negative returns tend to be larger in magnitude than positive returns, as investors in struggling economies or at-risk financial nodes often react more strongly to bad news than to good news, leading to asymmetric volatility \citep{barunik2016asymmetric}. 

\textcolor{blue}{Overall, two sources of asymmetry can be distinguished in the present framework. First, the volatility process itself can have asymmetric influences, where positive and negative shocks may have different impacts on subsequent and neighbouring volatility levels. Specifically, in this spatiotemporal setting, the conditional volatility at a location \( s_i \) depends not only on its own past information but also on contemporaneous shocks from neighbouring units:
\begin{eqnarray}
    \sigma_{t,s_i} = \sqrt{h_t(s_i)} = \sqrt{\text{Var}\!\left( Y_t(s_i) \,\middle|\, \mathcal{F}_{t-1,s_i} \cup \mathcal{F}_{t,-s_i} \right)},
\end{eqnarray}
where \( \mathcal{F}_{t-1,s_i} \) denotes the information set generated by past observations at location \( s_i \) and \( \mathcal{F}_{t,-s_i} \) represents the contemporaneous information from all other locations \( s_j \neq s_i \) at time \( t \). The inclusion of \( \mathcal{F}_{t,-s_i} \) in the conditioning set directly reflects the potential for instantaneous and directional spillovers.
This formulation departs fundamentally from classical GARCH models, where volatility is measurable solely with respect to the global past information set \( \mathcal{F}_{t-1} \). In the spatial-asymmetric framework, the conditioning set is inherently richer, and in general
$\sigma_{t,s_i} \neq \sqrt{\text{Var}\!\left( Y_t(s_i) \,\middle|\, \mathcal{F}_{t-1} \right)},$
as discussed in \cite{otto2021stochastic}. Second, the spatial interactions can be asymmetric, depending on the choice of the spatial weight matrix. While symmetric spatial weight matrices, such as inverse-distance weight matrices, offer interpretability and simplify inference, they may fail to capture the directional nature of volatility spillovers. Asymmetric matrices, by contrast, allow for non-reciprocal effects, where the influence from one location to another may not be matched in the reverse direction.}

 \subsection{Properties of the model}\label{sec:egarch_properties}

In this section, we analyse key properties of the model, including stationarity conditions and the explicit solution under which the process remains strictly and weakly stationary. Additionally, we derive the first two moments using the moment-generating function of the normal distribution. That is, we will later additionally assume normally distributed errors. While this serves as a natural starting point, extending the analysis to other error distributions, such as heavy-tailed distributions, is left for future research.

Assuming that the matrix $\xmat{I} - \lambda_0\xmat{W}_2$ has a full rank. From the definition $ \ln(\xvec{h_t})$ in (\ref{eq:unified2}), we get that:
\begin{eqnarray}\label{defF} \ln(\xvec{h_t}) & = & \lambda_1 (\xmat{I} - \lambda_0\xmat{W}_2)^{-1} \ln(\xvec{h_{t-1}}) + (\xmat{I} - \lambda_0\xmat{W}_2)^{-1}
\left(\xvec{\alpha}_1 + \rho_0 \xmat{W}_1 {g}(\xvec{\varepsilon}_t) +  \rho_1 g(\xvec{\varepsilon}_{t-1}) \right).
\end{eqnarray}
Thus, if considering the different locations as variables, $\ln(\xvec{h_t})$ can be written as a VAR(1) process: 
\begin{eqnarray}\label{defF2} \ln(\xvec{h_t}) & = & \lambda_1 (\xmat{I} - \lambda_0\xmat{W}_2)^{-1} \ln(\xvec{h_{t-1}}) + \xvec{\Delta}_t,
\end{eqnarray}
with time-dependent noise $\xvec{\Delta}_t$ given by: $\xvec{\Delta}_t = (\xmat{I} - \lambda_0\xmat{W}_2)^{-1}
\left(\xvec{\alpha}_1 + \rho_0 \xmat{W}_1 {g}(\xvec{\varepsilon}_t) +  \rho_1 g(\xvec{\varepsilon}_{t-1}) \right).$
As $\{ \xvec{\varepsilon}_t \}$ is an independent random sequence  with mean zero and existing covariance matrix  then it holds that $E(g(\xvec{\varepsilon}_t)) = {\bf 0}$, $E( \xvec{\Delta}_t ) = (\xmat{I} - \lambda_0\xmat{W}_2)^{-1}
\xvec{\alpha}_1$ and  
\[ Cov( \xvec{\Delta}_t ) =  (\xmat{I} - \lambda_0\xmat{W}_2)^{-1} \left( \rho_0^2 \xmat{W}_1 Cov(g(\xvec{\varepsilon}_t)) \xmat{W}_1^\prime + \rho_1^2 Cov(g(\xvec{\varepsilon}_{t-1})) \right) (\xmat{I} - \lambda_0\xmat{W}_2^\prime )^{-1} . \]
Determining \( \operatorname{Cov}(g(\xvec{\varepsilon}_t)) \) is not straightforward. However, if the covariance matrix of \( \xvec{\varepsilon}_t \) is diagonal, then \( \operatorname{Cov}(g(\xvec{\varepsilon}_t)) \) is also diagonal. Further,
\[ Var(g(\varepsilon_t(\xvec{s}_i))) = \Theta^2 Var(\varepsilon_1(\xvec{s}_i)) + \xi^2 Var(|\varepsilon_1(\xvec{s}_i)|) + 2 \Theta \xi E( \varepsilon_t(\xvec{s}_i ) | \varepsilon_t(\xvec{s}_i )| ) . \]
If, for example, $\varepsilon_t(\xvec{s}_i)$ follows a standard normal distribution, this expression simplifies to
\begin{equation}\label{varg} Var(g(\varepsilon_t(\xvec{s}_i))) = \Theta^2 + \xi^2 (1 - \frac{2}{\pi})   . \end{equation}
Moreover, note that $Cov(\xvec{\Delta}_t, \xvec{\Delta}_{t-1} ) \neq \xvec{0}$, while for $v\ge 2$, we have
$Cov(\xvec{\Delta}_t, \xvec{\Delta}_{t-v}) = \xvec{0}$. 


To analyse the properties of the spatiotemporal E-GARCH model, we establish conditions that ensure the process is well-defined and stationary. These conditions, which involve the parameters governing spatial dependence $\lambda_0$, temporal persistence $\lambda_1$, and the spatial weight matrix $\xmat{W}_2$, serve to prevent explosive behaviour in the volatility dynamics. Intuitively, they ensure that shocks do not propagate indefinitely across space and time, preserving the stability of the system. The following theorem formalises these conditions and provides an explicit solution for the process \( \{ \xvec{Y}_t \} \), along with its first two moments under the assumption of normally distributed innovations. Throughout, we use  $\odot$  to denote the Hadamard (element-wise) product.

\begin{thm}\label{th:existence}
\textcolor{blue}{	Suppose that 
 \begin{equation}\label{cond1}
\varrho(\lambda_1 (\xmat{I} - \lambda_0 \xmat{W}_2)^{-1}) < 1
\end{equation}
and 
\begin{equation}\label{cond2}
\varrho(\lambda_0 \xmat{W}_2) < 1 
\end{equation}
where $\varrho(\cdot)$ denotes the spectral radius of a matrix. Further let $\{ \xvec{\varepsilon}_t \}$ be a sequence of independent and identically distributed random vectors with mean zero and covariance matrix equal to the identity matrix then the process $\{ \xvec{Y}_t \}$ given by (\ref{eq:initial}), (\ref{eq:unified2}), and (\ref{defg}) has a unique strictly stationary solution given by} 
 \begin{equation}\label{eq:Y} \xvec{Y}_t = \xvec{\varepsilon}_t \; \odot \; \exp\left(\frac{1}{2} \sum_{v=0}^\infty \lambda_1^v (\xmat{I} - \lambda_0 \xmat{W}_2)^{-v} \xvec{\Delta}_{t-v} \right)  . \end{equation}
 Consequently,
 \[ Y_t(\xvec{s}_i) = \varepsilon_t(\xvec{s}_i) \; \exp\left(\frac{1}{2} \sum_{v=0}^\infty \lambda_1^v \xvec{e}_i^\prime (\xmat{I} - \lambda_0 \xmat{W}_2)^{-v} \xvec{\Delta}_{t-v} \right)  \]
 where $\xvec{e}_i$ denotes the n-dimensional vector whose i-th component is equal to $1$ and all others are equal to $0$.
 
 If further $\{ \xvec{\varepsilon}_t \}$ is an independent and  multivariate normally random process with mean $\xvec{0}$ and covariance matrix $\xmat{I}$ then the process $\{ \xvec{Y}_t \}$ given in (\ref{eq:Y}) is weakly stationary with
\begin{eqnarray*}
E( Y_t(\xvec{s}_i) ) & = & exp\left( \frac{1}{2} \xvec{e}_i^\prime \left((1-\lambda_1) \xmat{I} - \lambda_0 \xmat{W}_2\right)^{-1} \xvec{\alpha}_1\right) 
\\
& \quad \times & \; E\left( \varepsilon_1(\xvec{s}_i) \; exp\left( \frac{1}{2} \rho_0 \xvec{e}_i^\prime (\xmat{I} - \lambda_0 \xmat{W}_2)^{-1} \xmat{W}_1 g(\xvec{\varepsilon}_1) \right) \right)\\
& \quad \times & \prod_{v=1}^\infty \; E\left(exp\left(\frac{1}{2} \xvec{e}_i^\prime  \lambda_1^{v-1} (\xmat{I} - \lambda_0 \xmat{W}_2)^{-v} \left( \rho_0 \lambda_1 (\xmat{I} - \lambda_0 \xmat{W}_2)^{-1} \xmat{W}_1 + \rho_1 \xmat{I} \right) g(\xvec{\varepsilon}_{1}) \right)  \right),
\end{eqnarray*}

\begin{eqnarray*}
E( Y_t(\xvec{s}_i)^2 ) 
& = & exp\left(  \xvec{e}_i^\prime \left((1-\lambda_1) \xmat{I} - \lambda_0 \xmat{W}_2\right)^{-1} \xvec{\alpha}_1\right) \times \;
 E\left( \varepsilon_1(\xvec{s}_i)^2 \; exp\left( \rho_0 \xvec{e}_i^\prime (\xmat{I} - \lambda_0 \xmat{W}_2)^{-1} \xmat{W}_1 g(\xvec{\varepsilon}_1) \right) \right)\\
& \qquad \times & \prod_{v=1}^\infty  E\left(exp\left( \xvec{e}_i^\prime \lambda_1^{v-1} (\xmat{I} - \lambda_0 \xmat{W}_2)^{-v} \left( \rho_0 \lambda_1 (\xmat{I} - \lambda_0 \xmat{W}_2)^{-1} \xmat{W}_1 + \rho_1 \xmat{I} \right) g(\xvec{\varepsilon}_1)\right) \right)
\end{eqnarray*}

\begin{eqnarray*}
E( Y_t(\xvec{s}_i)  Y_t(\xvec{s}_j) ) & = & exp\left( \frac{1}{2} (\xvec{e}_i + \xvec{e}_j)^\prime \left((1-\lambda_1) \xmat{I} - \lambda_0 \xmat{W}_2\right)^{-1} \xvec{\alpha}_1\right)\\
& \qquad \times &  E\left( \varepsilon_1(\xvec{s}_i) \varepsilon_1(\xvec{s}_j) \; exp\left( \frac{1}{2} \; \rho_0 (\xvec{e}_i + \xvec{e}_j)^\prime (\xmat{I} - \lambda_0 \xmat{W}_2)^{-1} \xmat{W}_1 g(\xvec{\varepsilon}_1) \right) \right)\\
& \qquad \times & \prod_{v=1}^\infty  E\left(exp\left( \frac{1}{2} (\xvec{e}_i + \xvec{e}_j)^\prime \lambda_1^{v-1} (\xmat{I} - \lambda_0 \xmat{W}_2)^{-v} \left( \rho_0 \lambda_1 (\xmat{I} - \lambda_0 \xmat{W}_2)^{-1} \xmat{W}_1 + \rho_1 \xmat{I} \right) g(\xvec{\varepsilon}_{1})\right) \right)
\end{eqnarray*}

If $\xi = 0$ then 
\begin{eqnarray*}
E( Y_t(\xvec{s}_i) ) & = & exp\left( \frac{1}{2} \xvec{e}_i^\prime \left((1-\lambda_1) \xmat{I} - \lambda_0 \xmat{W}_2\right)^{-1} \xvec{\alpha}_1\right)\\
& \times&   \frac{1}{2} \Theta \rho_0 \xvec{e}_i^\prime (\xmat{I} - \lambda_0 \xmat{W}_2)^{-1} \xmat{W}_1 \xvec{e}_i \times \, \exp\left(\frac{1}{8} \rho_0^2 \Theta^2 \xvec{e}_i^\prime (\xmat{I} - \lambda_0 \xmat{W}_2)^{-1} \xmat{W}_1 \xmat{W}_1^\prime (\xmat{I} - \lambda_0 \xmat{W}_2^\prime)^{-1} \xvec{e}_i  \right)   \\
&   \times& \prod_{v=1}^\infty  \exp\left(\frac{1}{8} \Theta^2 \lambda_1^{2v-2} \xvec{e}_i^\prime  (\xmat{I} - \lambda_0 \xmat{W}_2)^{-v} \times \right. \\
& & \left. \times \left( \rho_0 \lambda_1 (\xmat{I} - \lambda_0 \xmat{W}_2)^{-1} \xmat{W}_1 + \rho_1 \xmat{I} \right) \left( \rho_0 \lambda_1 (\xmat{I} - \lambda_0 \xmat{W}_2)^{-1} \xmat{W}_1 + \rho_1 \xmat{I} \right)^\prime  
 (\xmat{I} - \lambda_0 \xmat{W}_2^\prime )^{-v} \xvec{e}_i 
\right) ,
\end{eqnarray*}
\begin{eqnarray*}
E( Y_t(\xvec{s}_i)^2 ) & = &  exp\left( \xvec{e}_i^\prime \left((1-\lambda_1) \xmat{I} - \lambda_0 \xmat{W}_2\right)^{-1} \xvec{\alpha}_1\right)\\
& & \times   \left(1 + \rho_0^2 \Theta^2 (\xvec{e}_i^\prime (\xmat{I} - \lambda_0 \xmat{W}_2)^{-1} \xmat{W}_1 \xvec{e}_i)^2 \right)\times \, \exp\left( \frac{1}{2} \rho_0^2 \Theta^2 \xvec{e}_i^\prime (\xmat{I} - \lambda_0 \xmat{W}_2)^{-1} \xmat{W}_1 \xmat{W}_1^\prime (\xmat{I} - \lambda_0 \xmat{W}_2^\prime)^{-1} \xvec{e}_i \right)   \\
&   & \times \prod_{v=1}^\infty  \exp\left(\frac{1}{2} \Theta^2 \lambda_1^{2v-2} \xvec{e}_i^\prime  (\xmat{I} - \lambda_0 \xmat{W}_2)^{-v} \times \right. \\
& & \times \left. \left( \rho_0 \lambda_1 (\xmat{I} - \lambda_0 \xmat{W}_2)^{-1} \xmat{W}_1 + \rho_1 \xmat{I} \right)  \left( \rho_0 \lambda_1 (\xmat{I} - \lambda_0 \xmat{W}_2)^{-1} \xmat{W}_1 + \rho_1 \xmat{I} \right)^\prime  (\xmat{I} - \lambda_0 \xmat{W}_2^\prime)^{-v} \xvec{e}_i \right)  .
\end{eqnarray*}
For $ i \neq j$ it holds that
\begin{eqnarray*}
E( Y_t(\xvec{s}_i)  Y_t(\xvec{s}_j) ) & = & exp\left( \frac{1}{2} (\xvec{e}_i + \xvec{e}_j)^\prime \left((1-\lambda_1) \xmat{I} - \lambda_0 \xmat{W}_2\right)^{-1} \xvec{\alpha}_1\right)\\
& &  \hspace*{-2cm} \times    \frac{1}{4} \Theta^2 \rho_0^2 (\xvec{e}_i + \xvec{e}_j)^\prime (\xmat{I} - \lambda_0 \xmat{W}_2)^{-1} \xmat{W}_1  \xvec{e}_i \times   (\xvec{e}_i + \xvec{e}_j)^\prime (\xmat{I} - \lambda_0 \xmat{W}_2)^{-1} \xmat{W}_1  \xvec{e}_j \\
& & \hspace*{-2cm} \times exp\left( \frac{1}{8} \rho_0^2 \Theta^2   (\xvec{e}_i + \xvec{e}_j)^\prime (\xmat{I} - \lambda_0 \xmat{W}_2)^{-1} \xmat{W}_1 \xmat{W}_1^\prime (\xmat{I} - \lambda_0 \xmat{W}_2^\prime)^{-1} (\xvec{e}_i + \xvec{e}_j) \right) \\
& & \hspace*{-2cm} \times \prod_{v=1}^\infty 
exp\left( \frac{1}{4} \Theta^2 \lambda_1^{2v-2} (\xvec{e}_i + \xvec{e}_j)^\prime (\xmat{I} - \lambda_0 \xmat{W}_2)^{-v} \left( \rho_0 \lambda_1 (\xmat{I} - \lambda_0 \xmat{W}_2)^{-1} \xmat{W}_1 + \rho_1 \xmat{I} \right) \times \right. \\
& & \hspace*{-0cm} \left. \times \left( \rho_0 \lambda_1 (\xmat{I} - \lambda_0 \xmat{W}_2)^{-1} \xmat{W}_1 + \rho_1 \xmat{I} \right)^\prime (\xmat{I} - \lambda_0 \xmat{W}_2^\prime)^{-v}
(\xvec{e}_i + \xvec{e}_j) \right) .
\end{eqnarray*}

\end{thm}

\textcolor{blue}{The proof of this theorem is given in the Appendix. Note that conditions provided in~\eqref{cond1} and~\eqref{cond2} involve assumptions on the eigenvalues of certain matrices, which cannot be easily verified in practice. Using the fact that $\varrho(\xmat{A}) \leq \|\xmat{A}\|$ for any submultiplicative matrix norm, and replacing $\varrho(\cdot)$ by $\|\cdot\|$ in these conditions, we obtain stronger but more easily verifiable conditions. In particular, condition~\eqref{cond2} implies}                       
\[ \|(\xmat{I} - \lambda_0 \xmat{W}_2)^{-1}\| = \left\| \sum_{v=0}^\infty \lambda_0^v \xmat{W}_2^v \right\| \leq \sum_{v=0}^\infty |\lambda_0^v| \|\xmat{W}_2^v \| = \frac{1}{1-|\lambda_0| \|\xmat{W}_2\|} . \]
Consequently, the assumptions \eqref{cond1} and \eqref{cond2} are fulfilled if $|\lambda_1| + |\lambda_0| \; \|\xmat{W}_2\| < 1$. In practice, spatial weight matrices are often normalised, e.g., row-standardised, i.e., the sum of the column elements is equal to one for all rows. Then, choosing the $\|.\|_\infty$ matrix norm, the above condition reduces to $|\lambda_0| + |\lambda_1| < 1$ \textcolor{blue}{since for a row-standardised matrix $\|\xmat{W}_2\|_\infty = 1$. This property follows from the definition of the $\|\cdot\|_\infty$ norm as the maximum absolute row sum, which for a row-standardised matrix equals one.}.  

Moreover, applying a log-squared transformation of the observations (\citealt{robinson2009large}), the spatiotemporal E-GARCH model can be rewritten as a spatiotemporal autoregressive process with lagged values of innovations. Thus, this model aligns with a spatiotemporal autoregressive process applied to the log-squared transformation of the process $\ln \xvec{Y}_t^2$. The estimation of parameters can be carried out as detailed in \cite{sato2017spatial}. It is a special VAR(1) process with time-correlated noise.

The following theorem formalises this result and establishes the necessary conditions to derive the key properties of the transformed model. In particular, it characterises the mean and covariance structure of the resulting process, further demonstrating how the dependence parameters influence the spatiotemporal dynamics of volatility. 

\begin{thm}\label{theorem2}
Suppose that the assumptions (\ref{cond1}) and (\ref{cond2}) are fulfilled. If further $\{ \xvec{\varepsilon}_t \}$ is an independent and  multivariate normally distributed random process with mean $\xvec{0}$ and covariance matrix $\xmat{I}$. With,
 \begin{eqnarray}
	\xvec{\nu}_t & = &  \ln \xvec{Y}_t^2 - (\xmat{I} - \lambda_0 \xmat{W}_2)^{-1} \lambda_1\ln \xvec{Y}_{t-1}^2 \nonumber \\
	& = &  (\xmat{I} - \lambda_0 \xmat{W}_2)^{-1} (\xvec{\alpha}_1 + \rho_0 \xmat{W}_1 g(\xvec{\varepsilon}_t) + \rho_1 g(\xvec{\varepsilon}_{t-1}) {\color{blue} - } \lambda_1 \ln \xvec{\varepsilon}_{t-1}^2) + \ln \xvec{\varepsilon}_t^2\\
 & = & \xmat{\Delta}_t - \lambda_1 (\xmat{I} - \lambda_0 \xmat{W}_2)^{-1} \ln \xvec{\varepsilon}_{t-1}^2 + \ln \xvec{\varepsilon}_t^2 \, ,
\end{eqnarray} 
the model can be rewritten as 
 \begin{eqnarray}\label{eqcontinu}
\ln \xvec{Y}_t^2	 & = &    (\xmat{I} - \lambda_0 \xmat{W}_2)^{-1} \lambda_1\ln \xvec{Y}_{t-1}^2+\xvec{\nu}_t .
\end{eqnarray} and it holds that:
\begin{itemize}
    \item[a)]
\begin{eqnarray*}
	E(\xvec{\nu}_t) 
 & = &  (\xmat{I} - \lambda_0 \xmat{W}_2)^{-1}\xvec{\alpha}_1  + (-\ln(2) - \gamma ) (\xmat{I} - \lambda_1 (\xmat{I} - \lambda_0 \xmat{W}_2)^{-1}) \xvec{1} 
\end{eqnarray*}
where $\gamma \approx 0.57721...$ stands for the Euler-Mascheroni constant. If further $\xmat{W}_2$ is row-standardized then 
\begin{eqnarray*}
	E(\xvec{\nu}_t) 
 & = &  (\xmat{I} - \lambda_0 \xmat{W}_2)^{-1}\xvec{\alpha}_1  + (-\ln(2) - \gamma ) 
 (1-\frac{\lambda_1}{1-\lambda_0}) \xvec{1} .  
\end{eqnarray*}
\item[b)] 
\begin{eqnarray*}
   Cov(\xvec{\nu}_t) & = &  \rho_0^2 (\Theta^2 + \xi^2(1-\frac{2}{\pi})) (\xmat{I} - \lambda_0 \xmat{W}_2)^{-1} \xmat{W}_1 \xmat{W}_1^\prime(\xmat{I} - \lambda_0 \xmat{W}_2^\prime)^{-1} + \frac{\pi^2}{2} \; \xmat{I} \\
& & + \rho_0  2 \xi \ln(2) \sqrt{\frac{2}{\pi}} \left( (\xmat{I} - \lambda_0 \xmat{W}_2)^{-1} \xmat{W}_1 + \xmat{W}_1^\prime(\xmat{I} - \lambda_0 \xmat{W}_2^\prime)^{-1} \right) \\
& & + \left(\rho_1^2 (\Theta^2 + \xi^2 (1- \frac{2}{\pi})) + \lambda_1^2 \frac{\pi^2}{2} - 4 \ln(2) \rho_1 \lambda_1 \xi \sqrt{\frac{2}{\pi}}\right) 
   (\xmat{I} - \lambda_0 \xmat{W}_2)^{-1}(\xmat{I} - \lambda_0 \xmat{W}_2^\prime)^{-1} .\end{eqnarray*}
\item[c)] 
\begin{eqnarray*}
    Cov(\xvec{\nu}_t, \xvec{\nu}_{t-1}) & = & ( \rho_0 \rho_1 (\Theta^2 + \xi^2 (1 - \frac{2}{\pi})) - 2 \lambda_1 \rho_0 \xi \ln(2) \sqrt{\frac{2}{\pi}}) \; (\xmat{I} - \lambda_0 \xmat{W}_2)^{-1} \xmat{W}_1^\prime (\xmat{I} - \lambda_0 \xmat{W}_2^\prime)^{-1}\\
& & + (2 \rho_1 \xi \ln(2) \sqrt{\frac{2}{\pi}} - \lambda_1 \frac{\pi^2}{2}) 
\; (\xmat{I} - \lambda_0 \xmat{W}_2)^{-1} . \end{eqnarray*} 
\item[d)]  $Cov(\xvec{\nu}_t, \xvec{\nu}_{t-s}) = 0$ for $s > 1$.
\end{itemize}   
\end{thm}

The results derived in Theorem \ref{theorem2} could serve as a foundation for developing a weighted least squares (WLS) estimation procedure, leveraging the structure of the transformed process. However, exploring such an approach is beyond the scope of this paper and remains a direction for future research. Instead, in the following section, we focus on parameter estimation based on the maximum likelihood principle, which provides a well-established and flexible framework for inference in spatiotemporal GARCH models.

\subsection{Parameter estimation}\label{sec:egarch_inference}

In the following section, we focus on estimating the parameters  $\vartheta = (\alpha, \rho_0, \rho_1, \lambda_0, \lambda_1, \Theta, \xi)'$, where, for identification purposes, we set $\xi=1$. A key challenge in the maximum likelihood estimation arises from the fact that the mapping from the observed values \( \xvec{Y}_t \) to the underlying innovations \( \xvec{\varepsilon}_t \) is not available in closed form, because log-volatility term also depends on the contemporaneous innovations $\xvec{\varepsilon}_t$ for the spatial effects (see also above for details). Therefore, we first examine this inversion step in detail, analysing its implications for estimation. Based on these insights, we then propose a quasi-maximum likelihood (QML) estimator, which leverages the structure of the model for practical inference while accounting for the required numerical inversion.



The maximum-likelihood principle is based on considering the joint density of the present data $\xvec{Y}_1, \cdots , \xvec{Y}_T$. This quantity is difficult to determine since $\xvec{Y}_t$ depends on all $\xvec{\varepsilon}_v, v \le t$. Unfortunately, the mapping from $(\xvec{\varepsilon}_1, \cdots, \xvec{\varepsilon}_T)$ to $(\xvec{Y}_1, \cdots,\xvec{Y}_T)$ is not injective and, thus, a unique inverse function does not exist. To address this, we propose a quasi-maximum likelihood (QML) approach, which employs numerical inversion, assuming that the initial observations of the time series are known.


In the next theorem, it is shown that if $\xvec{Y}_0$ and $\xvec{\varepsilon}_0$ are known then $\xvec{Y}_t = f_t(\xvec{\varepsilon}_1, \cdots, \xvec{\varepsilon}_t), t=1, \cdots ,T$ and the inverse of the function 
\[ ( \xvec{Y}_t )_{t=1,\cdots,T} = ( f_t(\xvec{\varepsilon}_1, \cdots , \xvec{\varepsilon}_t) )_{t=1, \cdots ,T} = {\xvec{f}}_T(\xvec{\varepsilon}_1, \cdots , \xvec{\varepsilon}_T)  \] 
exists. Thus it is possible to apply the transformation rule for random vectors to determine the likelihood function of $\xvec{Y}_1, \cdots, \xvec{Y}_T$. In practice, we suggest that the first observations of the time series are discarded for estimation until the influence of $\xvec{Y}_0$ and $\xvec{\varepsilon}_0$ vanishes, as we will illustrate in the following Section \ref{sec:MC}.

\begin{thm}\label{th:inversion}
	Let  $T \ge 1$. Suppose that the assumptions of Theorem \ref{th:existence} are satisfied and suppose that $\xvec{Y}_0$ and $\xvec{\varepsilon}_0$ are known values. Assume that$\prod_{v=1}^T \prod_{i=1}^n \xvec{\varepsilon}_v(\xvec{s}_i) \neq 0$ almost surely. The inverse of the function $( \xvec{Y}_{t} )_{t=1,\cdots,T} = \xvec{f}_T(\xvec{\varepsilon}_{1},\cdots,\xvec{\varepsilon}_{T})$ exists if for all $1 \le v \le T$, the following condition hold almost surely:
 \begin{eqnarray}
     \det\left( \xmat{I} + \frac{1}{2} \rho_0 (\xmat{I} - \lambda_0 \xmat{W}_2)^{-1} \xmat{W}_1 \odot ( \Theta \xvec{\varepsilon}_v \xvec{1}^\prime + \xi \xvec{\varepsilon}_v  sgn(\xvec{\varepsilon}_{v-1})^\prime )\right) \neq 0 . 
 \end{eqnarray}
\end{thm}

Now, it is possible to calculate the conditional likelihood function of $\xvec{Y}_1, \cdots, \xvec{Y}_T$ given $\xvec{Y}_0$ and $\xvec{\varepsilon}_0$. Note that $\xvec{Y}_t = f_t(\xvec{\varepsilon}_1,..., \xvec{\varepsilon}_t), t=1,...,T$ and 
\[ \xvec{\varepsilon}_t = g_t(\xvec{Y}_1,..., \xvec{Y}_t) = g_t(\xvec{Y}^*_t), t=1,...,T \] 
with $\xvec{Y}_t^* = (\xvec{Y}_1^\prime,..., \xvec{Y}_t^\prime)^\prime$ as shown above. It is given by 
\begin{eqnarray}
    l(\vartheta, \xvec{y}_1, \xvec{y}_2, \cdots, \xvec{y}_T| \xvec{\varepsilon}_0,  \xvec{y}_0 ) 
    & = & \frac{1}{|\det(J_{\xvec{y}})|} \;  \prod_{t=1}^T f_{\xvec{\varepsilon}}( g_t(\xvec{y}_t^*))     
\end{eqnarray}
Here $f_{\xvec{\varepsilon}}$ stands for the density of the n-variate standard normal distribution and $J_{\xvec{y}}$ for the Jacobian matrix determined at the value $(g_1(\xvec{y}_1^*), \ldots , g_T(\xvec{y}_T^*))$, i.e., $J_{\xvec{y}} = J_{\xvec{\varepsilon}}(g_1(\xvec{y}_1^*), \ldots , g_T(\xvec{y}_T^*))$ with  
\[ J_{\xvec{\varepsilon}} = \left( \frac{\partial \xvec{Y}_i}{\partial \xvec{\varepsilon}_j} \right)_{\begin{array}{l} i=1,..,T\\[-0.2cm]
 j=1,..,T \end{array}}  . \]
Since $\frac{\partial \xvec{Y}_i}{\partial \xvec{\varepsilon}_j} = \xmat{0}$ for $j >i$ the determinant of $J_{\xvec{\varepsilon}}$ is equal to the product of the determinants of 
$J_t =\frac{\partial \xvec{Y}_t}{\partial \xvec{\varepsilon}_t} $ for $t=1,..,T$. Consequently,
\begin{eqnarray}
    l(\vartheta, \xvec{y}_1, \xvec{y}_2, \cdots, \xvec{y}_T| \xvec{\varepsilon}_0,  \xvec{y}_0 ) 
    & = & \prod_{t=1}^T \frac{1}{|\det(J_t(g_1(\xvec{y}_1^*),..., g_T(\xvec{y}_T^*))  |} \;  \prod_{t=1}^T f_{\xvec{\varepsilon}}( g_t(\xvec{y}_t^*))     
\end{eqnarray}
Now, 
\begin{eqnarray}
    J_t = \left(
  \begin{array}{cccccc}
    \sqrt{h_t(s_1)} + \frac{\partial \sqrt{h_t(\xvec{s}_1)} }{\partial \varepsilon_t(\xvec{s}_1)}\varepsilon_t(\xvec{s}_{1}) &   \frac{\partial \sqrt{h_t(\xvec{s}_1)} }{\partial \varepsilon_t(\xvec{s}_2)}\varepsilon_t(\xvec{s}_{1}) &  & \cdots &  & \frac{\partial \sqrt{h_t(\xvec{s}_1)} }{\partial \varepsilon_t(\xvec{s}_n)}\varepsilon_t(\xvec{s}_1) \\
\\
     \frac{\partial \sqrt{h_t(\xvec{s}_2)} }{\partial \varepsilon_t(\xvec{s}_1)}\varepsilon_t(\xvec{s}_{2}) & &  & \cdots &  & \frac{\partial \sqrt{h_t(\xvec{s}_2)} }{\partial \varepsilon_t(\xvec{s}_n)}\varepsilon_t(\xvec{s}_2) \\

     &  &  &  &  &  \\
    \vdots &  &  & \ddots &  & \vdots \\

     &  &  &  &  &  \\
     \frac{\partial \sqrt{h_t(\xvec{s}_{n-1})} }{\partial \varepsilon_t(\xvec{s}_1)} \varepsilon_t(\xvec{s}_{n-1}) & &  & \cdots &  & \frac{\partial \sqrt{h_t(\xvec{s}_{n-1})} }{\partial \varepsilon_t(\xvec{s}_n)}\varepsilon_t(\xvec{s}_{n-1}) \\
   \\
  \frac{\partial \sqrt{h_t(\xvec{s}_n)} }{\partial \varepsilon_t(\xvec{s}_1)}\varepsilon_t(\xvec{s}_n) & \cdots &   &  &   \frac{\partial \sqrt{h_t(\xvec{s}_n)} }{\partial \varepsilon_t(\xvec{s}_{n-1})}\varepsilon_t(\xvec{s}_{n}) &  \sqrt{h_t(\xvec{s}_n)} + \frac{\partial \sqrt{h_t(\xvec{s}_n)} }{\partial \varepsilon_t(\xvec{s}_n)}\varepsilon_t(\xvec{s}_n) \end{array} \right)
\end{eqnarray}
and
\[ \frac{\sqrt{h_t(\xvec{s}_i)} }{\partial \varepsilon_t(\xvec{s}_j)} = \frac{1}{2} \sqrt{h_t(\xvec{s}_i)} \frac{\partial{\ln h_t(\xvec{s}_i)} }{\partial \varepsilon_t(\xvec{s}_j)} = \frac{1}{2} \; \sqrt{h_t(\xvec{s}_i)} \; b_{ij} \;  (\Theta  + \xi sgn(\varepsilon_t(\xvec{s}_j)) ) . \]

A key question is how the function $g_t$ can be determined in practice. Applying the log-square transformation to the observations \citep{robinson2009large}, we obtain
\[ 
\ln( \xvec{Y}_t^2 ) = \ln( \xvec{h}_t ) + \ln( \xvec{\varepsilon}_t^2 ) . \]
Using \eqref{eq:unified2}, it follows that
\begin{equation}\label{eq:inversion}
(\xmat{I} - \lambda_0 \xmat{W}_2) ( \ln( \xvec{Y}_t^2 ) - \ln( \xvec{\varepsilon}_t^2 ) ) = \xvec{\alpha}_1 + \rho_0 \xmat{W}_1 g(\xvec{\varepsilon}_t) + \rho_1 g(\xvec{\varepsilon}_{t-1}) + \lambda_1 ( \ln( \xvec{Y}_{t-1}^2 ) - \ln( \xvec{\varepsilon}_{t-1}^2 ) ), t \ge 1 .
\end{equation}
Assuming $\xvec{\varepsilon}_0$ and $\xvec{Y}_0$ to be known, the model can be expressed as a stochastic difference equation, which can be inverted using a multivariate Newton procedure \citep[cf.][]{dennis1996numerical}. To solve this system of non-linear equations, we utilise the R-package \texttt{nleqslv} \citep{hasselman2018package}.  The sign of $\xvec{\varepsilon}_t$ is obtained by using the information that it has the same sign as $\xvec{Y}_t$. Consequently we put $\xvec{\varepsilon}_t = |\xvec{\varepsilon}_t|  sign(\xvec{Y}_t)$, $t=1,...,T $.

To assess the statistical uncertainty of the parameter estimates, we propose obtaining standard errors from the Hessian matrix of the log-likelihood function, which can computed numerically. Specifically, let $l(\vartheta, \xvec{y}_1, \xvec{y}_2, \cdots, \xvec{y}_T| \xvec{\varepsilon}_0,  \xvec{y}_0)$  denote the log-likelihood function evaluated at the parameter vector $\vartheta$. The observed information matrix is then given by the Hessian matrix
\[
\hat{\mathcal{I}}(\vartheta) = - \frac{\partial^2 l(\vartheta, \xvec{y}_1, \xvec{y}_2, \cdots, \xvec{y}_T| \xvec{\varepsilon}_0,  \xvec{y}_0 )}{\partial \vartheta \, \partial \vartheta^{\prime}}
\]
computed at the estimated parameter $\hat{\vartheta}$. The standard errors of the estimated parameters are obtained as the square roots of the diagonal elements of the inverse Hessian matrix, i.e.,
\[
\operatorname{SE}(\hat{\vartheta}) = \sqrt{\operatorname{diag}(\hat{\mathcal{I}}(\hat{\vartheta})^{-1})}.
\]
Since the Hessian is computed numerically, ensuring its invertibility and numerical stability is crucial. In cases where the Hessian is near-singular, regularisation techniques or alternative methods for standard error estimation may be required.

\section{Simulation studies}\label{sec:applications}

In the following section, we conduct simulation studies to evaluate the performance of the proposed estimation method. In particular, we analyse the numerical inversion procedure and assess the finite-sample accuracy of the QMLE under different settings.

\subsection{Numerical invertibility}\label{sec:MC}

In practical applications, the innovations \( \xvec{\varepsilon}_t \) are unobservable, and only \( \xvec{Y}_t \) is available. This poses a challenge in E-GARCH models, which becomes even more pronounced in the spatiotemporal setting due to the presence of contemporaneous innovations within \( \xvec{Y}_t \). Since the likelihood function is defined in terms of the innovations, estimating the parameters requires recovering \( \xvec{\varepsilon}_t \) numerically. In this section, we outline the numerical procedure employed in our simulations to invert the process and evaluate the accuracy of this numerical procedure.

We considered regular spatial grids of $n = 25$ locations and $T = 50$ time points. The spatial weights matrices, $\xmat{W}_1$ and $\xmat{W}_2$ were chosen using the Queen's and Rook's contiguity criterion respectively. It should be noted that the choice of the spatial weight matrix can influence the consistency of the estimation. The Rooks contiguity matrix is usually sparse and takes into account the immediate neighbours of each spatial unit, which can lead to more localised interactions being captured in the model. In contrast, the Queen's matrix considers all surrounding neighbours thus potentially leading to denser and more widespread interactions. The matrices have zeros on the diagonal and are row-standardised.

Moreover, for the classical E-GARCH model in time series settings, \cite{francq2019garch} suggested setting \(\xi\) equal to 1, at least for a time lag of 1, to avoid identification issues. Following this recommendation, we set \(\xi = 1\) in our simulations. We considered $\vartheta_0 = (\rho_0 = 0.25, \rho_1 = 0.3, \theta = 0.4, \alpha = 0.5,  \lambda_1 = 0.4, \lambda_0 = 0.35)'$ as the data-generating parameter set, and \(\xvec{\varepsilon}_t\) was generated from the standard normal distribution. We define the innovation estimate obtained from the inversion process as $\tilde{\xvec{\varepsilon}}_{t, \vartheta}$, which is computed recursively based on solving \eqref{eq:inversion} with respect to $\tilde{\xvec{\varepsilon}}_{t, \vartheta}$ for a given set of parameters $\vartheta$, i.e.,
\begin{eqnarray}\label{eq:inversion2}
    & (\xvec{I}& - \lambda_0 \xvec{W}_2)  \ln  \tilde{\xvec{\varepsilon}}_{t, \vartheta}^2 + \rho_0 \xmat{W}_1 g( \tilde{\xvec{\varepsilon}}_{t, \vartheta}) \; - \nonumber \\
    & \quad &\left((\xvec{I} - \lambda_0 \xvec{W}_2) \ln \xvec{Y}_t^2 - \xvec{\alpha}_1 - \rho_1 g( \tilde{\xvec{\varepsilon}}_{t-1, \vartheta}) - \lambda_1 \ln \xvec{Y}_{t-1}^2 + \lambda_1 \ln  \tilde{\xvec{\varepsilon}}_{t-1, \vartheta}^2 \right) = 0.
\end{eqnarray}
We considered initial values $\xvec{Y}_0(s) = 0.0001 $ and $\xvec{\varepsilon}_0(s) = 0.0001 $ for all locations.
We also assumed that $E|\xvec{\varepsilon}_t|$ is constant and equal to the expectation of the standard normal distribution. For each $t$, the system of non-linear equations is solved using the \texttt{nleqslv} solver, see above for further details. It is worth noting that this numerical inversion needs to be repeated for each parameter set $\vartheta$ inside the numerical maximisation of the log-likelihood. Thus, an efficient solver such as the \texttt{nleqslv} solver is important for a good computational performance. Below, we first focus on the differences between $\tilde{\xvec{\varepsilon}}_t$ and $\xvec{\varepsilon}_t$
for the true data-generating parameters $\vartheta_0$ for each time point $t$. Secondly, we analyse the same differences but for parameters $\vartheta$ away from  $\vartheta_0$ to show that the differences are, indeed, minimised at $\vartheta = \vartheta_0$.

\begin{figure}
  \centering
 \includegraphics[width=0.65\textwidth]{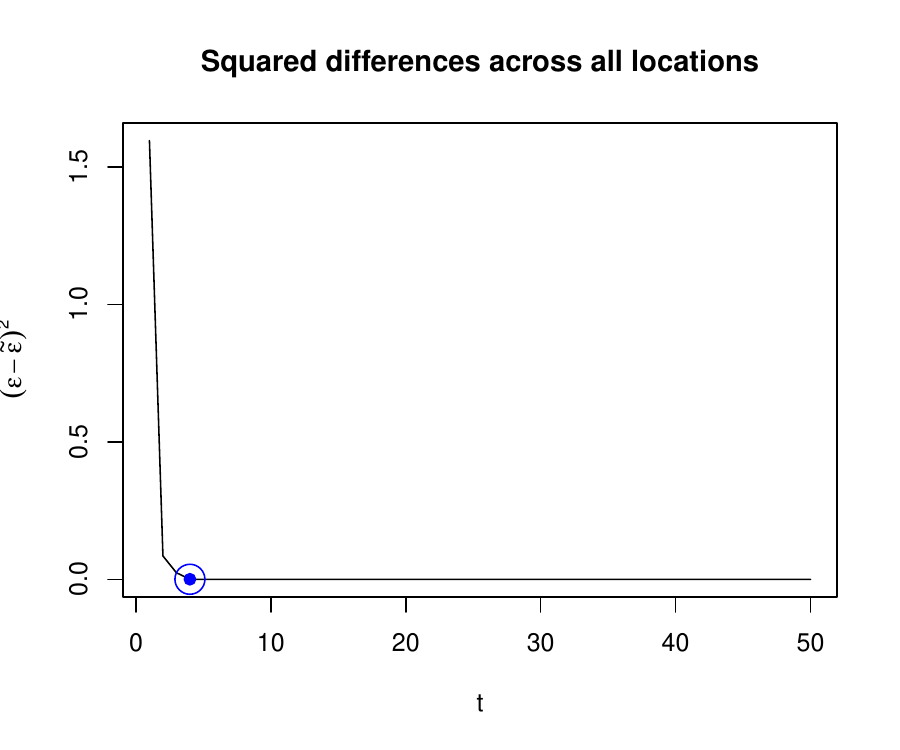}
  \caption{Maximum squared differences $\text{MaxD}_{\vartheta_0, t}$.}\label{fig:squareddiff}
\end{figure}

Figure \ref{fig:squareddiff} illustrates the maximum squared difference across all locations, that is,
\begin{equation}
    \text{MaxD}_{\vartheta_0, t} = \max_{i \in \{1, \ldots, n\}} ( \tilde{\varepsilon}_{t, \vartheta}(\xvec{s}_i) - \varepsilon_t(\xvec{s}_i))^2 \, .
\end{equation} 
The plot shows that after approximately five time steps, the maximum squared differences stabilised at values close to zero, indicating a high accuracy of the numerical inversion. The initial discrepancies in the first few observations arise due to conditioning on the initial values, which affects the early iterations of the inversion process. Therefore, we recommend discarding the first five values when computing the log-likelihood to ensure robust estimation.

\begin{figure}
  \centering
 \includegraphics[width=\textwidth]{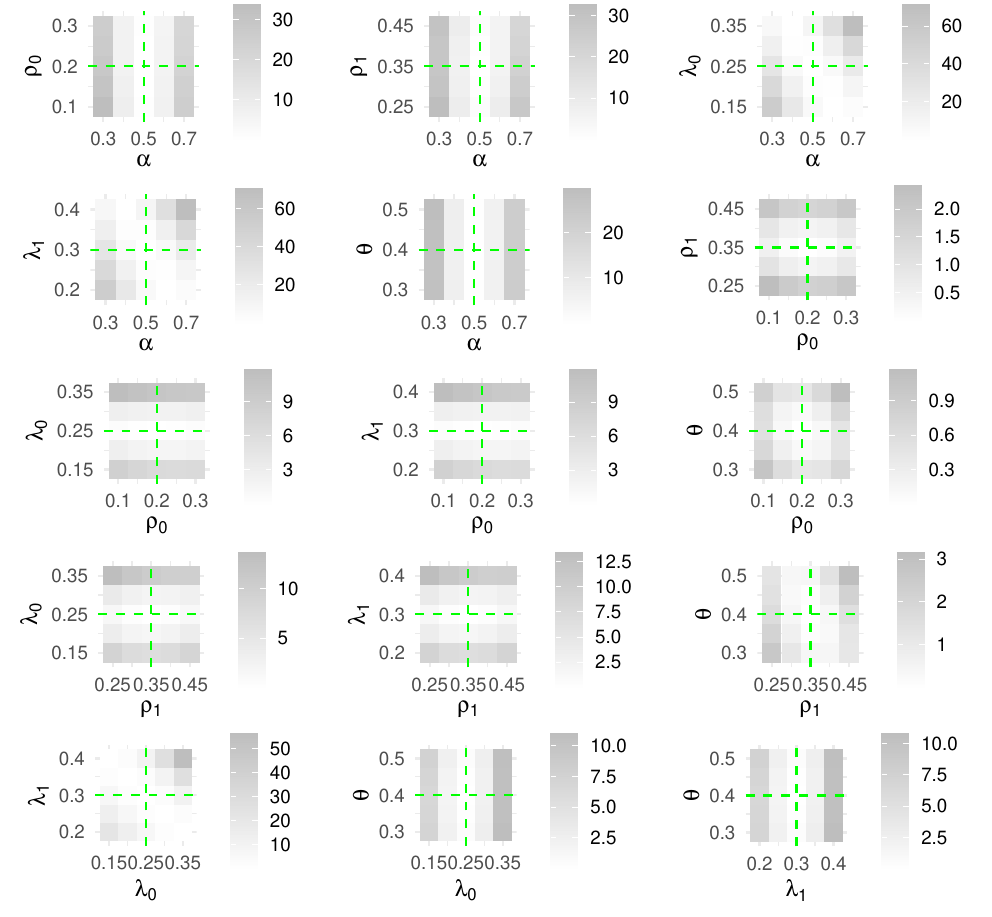}
 \caption{$\text{SSD}_{\vartheta}$ for different values of $\vartheta$ around the true data-generating $\vartheta_0$, indicated as vertical/horizontal green lines. The heatmaps are displayed for different pairs of parameters, where all remaining parameters are fixed at their true data-generating values.}\label{fig:figure2}
\end{figure}

Figure \ref{fig:figure2} displays the mean error of the inversion process over the 50 replications, with the first five values removed to avoid initialisation effects. The results illustrate that the reconstructed innovations \( \xvec{\tilde{\varepsilon}}_t \) align more closely with the true innovations \( \xvec{\varepsilon}_t \) when the parameters are near their true values.

As mentioned above, the inversion mapping depends on $\vartheta$. To understand how different parameter choices affect this process, we measure the sum squared differences discarding the first five observations, 
\begin{equation}
    \text{SSD}_{\vartheta} = \sum_{i = 1}^{n}\sum_{t = 6}^{T} ( \tilde{\varepsilon}_{t, \vartheta}(\xvec{s}_i) - \varepsilon_t(\xvec{s}_i))^2 \, ,
\end{equation} 
for a range of parameters 
\begin{equation}
    \vartheta = (\alpha, \rho_0, \rho_1, \lambda_0, \lambda_1, \theta)' \in [0.3, 0.7] \times [0.1, 0.3] \times [0.25, 0.45] \times [0.15, 0.35] \times [0.2, 0.4].
\end{equation}
The resulting 6-dimensional space is visualised by pairing all  parameters and showing the $\text{SSD}_{\vartheta}$ as a heat map, while keeping all remaining four parameters fixed at their true values. This simulation was repeated for 50 replications and the resulting average $\text{SSD}_{\vartheta}$ are shown in Figure \ref{fig:figure2}. The data-generating parameter set $\vartheta_0$ is indicated by the green cross in the centre. The differences $\text{SSD}_{\vartheta}$ are jointly minimised in the centre, i.e., at $\vartheta_0$, where the minimum is close to zero, indicating that the inversion performs best when the parameters are equal to their true data-generating values. Examining the range of these plots, we observe that some parameters lead to larger average difference, e.g. the constant volatility term $\alpha$ and and the GARCH coefficients $\lambda_0$ and $\lambda_1$, compared to other parameter combinations, such as those involving the asymmetry  parameter $\theta$.


\subsection{Maximum likelihood estimation results}

To assess the \textcolor{blue}{performance} of the QMLE for finite samples, we performed a Monte-Carlo simulation study for (a) an increasing size of 2-dimensional spatial unit grids, $4 \times 4$ ($n = 16$), $5 \times 5$ ($n = 25$) and $7 \times 7$ ($n = 49$), and (b) an increasing time horizon, varying from $T = 50$, $T = 100$ to $T = 150$. The weight matrices are chosen as in the examples above, i.e., we utilised row-normalised Queen's contiguity spatial weights matrices for $W_1$, and Rook's contiguity for $W_2$. 
The innovation terms $\xvec{\varepsilon}_t$ were simulated independently from a standard normal distribution, i.e.,  $\xvec{\varepsilon}_t \sim \mathcal{N}(0, 1)$. 

Moreover, we considered different models with $\vartheta_0 = \left(0.5, 0.35, 0.4, 0.5, 0.3, 0.2\right)'$ for model A with a pronounced contemporaneous spatial dependence and $\vartheta_0 = \left(0.2, 0.35, 0.4, 0.5, 0.3, 0.25\right)$  for model B, where the temporal effects dominate the spatial ones. 
The first five time points are excluded in the implementation of the quasi-maximum likelihood estimation to eliminate potential initialisation biases and ensure more reliable parameter estimates \ref{fig:squareddiff}. It is worth noting that the log-likelihood function is relatively flat---as it is typical for GARCH-type models---making it hard to find the optimum, thus the estimated parameters may highly depend on the initial values. The optimizer’s starting values can be randomised, and its step size and stopping criterion can be adjusted to improve convergence and robustness.
The results, presented in Table \ref{tab:parameters}, are based on $m = 1000$ replications reporting the Monte Carlo mean biases
\begin{equation}
    \text{Bias}_i = \frac{1}{m} \sum_{k = 1}^{m} (\hat{\vartheta}_i - \vartheta_i)
\end{equation}
and the root means squared errors
\begin{equation}
    \text{RMSE}_i = \sqrt{\frac{1}{m} \sum_{k = 1}^{m} (\hat{\vartheta}_i - \vartheta_i)^2}
\end{equation}
for each parameter from $i = 1, \ldots, 6$. 
Across both settings, bias values are generally small and fluctuate around zero, indicating that the estimator provides reasonably accurate estimates. Moreover, the RMSE decreases with increasing $T$ or increasing $n$. Additionally, the bias and RMSE tend to be larger for the spatial dependence parameters $\rho_0 ,  \lambda_0$ compared to other parameters, indicating greater sensitivity in estimating these effects.

The simulations were performed on a machine with four Intel Xeon Platinum 8276L CPUs (each with 28 cores, totalling 224 logical cores) and 1TB of RAM. For an average of 1000 replications, the computation time per estimation of one iteration ranges from 7.36 seconds for the smallest setting ($n=16$ and $T=50$) and 17.52 seconds for the largest setting ($n=49$ and $T=150$).

\begin{landscape}
\begin{table}
  \centering
  \caption{Bias and RMSE of the QMLE for model A and B.}\label{tab:parameters}
  \begin{tabular}{cccccccc|cccccc}
  \hline
  \multirow{2}{*}{} &   & \multicolumn{6}{c|}{Bias} & \multicolumn{6}{c}{RMSE} \\
   \hline
    \multicolumn{14}{l}{Model A} \\
    \hline
    $T$ & $n$ & $\rho_0=0.5$ & $\rho_1=0.35$ & $\theta=0.4$ & $\alpha=0.5$ & $\lambda_1=0.3$ & $\lambda_0=0.2$ & $\rho_0=0.5$ & $\rho_1=0.35$ & $\theta=0.4$ & $\alpha=0.5$ & $\lambda_1=0.3$ & $\lambda_0=0.2$ \\
    \hline
    &  & & & & & & 
     \\
    50 & 16 & -0.011 & -0.010 & 0.031 & 0.045 & -0.037 & -0.004 & 0.172 & 0.089 & 0.175 & 0.238 & 0.169 & 0.194 \\
     & 25 &  -0.001 & -0.005 & 0.014 & 0.044 & -0.028 & -0.012 & 0.144 & 0.071 & 0.133 & 0.202 & 0.139 & 0.163  \\
     & 49 & 0.006 & 0.001 & 0.002 & 0.038 & -0.021 & -0.015 & 0.108 & 0.049 & 0.090 & 0.162 & 0.108 & 0.134\\
     &  & & & & & & 
     \\
    
    100 & 16 & 0.002 & -0.002 & 0.007 & 0.039 & -0.012 & -0.024  & 0.117 & 0.061 & 0.114 & 0.169 & 0.118 & 0.144   \\
     & 25 & 0.002 & 0.000 & 0.003 & 0.028 & -0.011 & -0.016  & 0.095 & 0.050 & 0.088 & 0.151 & 0.101 & 0.122  \\
       & 49 & 0.007 & 0.002 & 0.002 & 0.032 & -0.010 & -0.019  & 0.072 & 0.034 & 0.062 & 0.123 & 0.078 & 0.099   \\
        &  & & & & & & 
     \\
    150 & 16 & 0.004 & 0.002 & 0.003 & 0.038 & -0.019 & -0.017  & 0.095 & 0.050 & 0.085 & 0.148 & 0.104 & 0.125  \\
      & 25 &  0.000 & 0.001 & 0.004 & 0.025 & -0.011 & -0.012  & 0.076 & 0.040 & 0.068 & 0.121 & 0.082 & 0.110 \\
       & 49 &   0.010 & 0.006 & -0.003 & 0.043 & -0.020 & -0.021 & 0.062 & 0.032 & 0.054 & 0.135 & 0.082 & 0.087 \\
        &  & & & & & & 
     \\
    \hline
    \multicolumn{14}{l}{Model B} \\
    \hline
    $T$ & $n$ & $\rho_0=0.2$ & $\rho_1=0.35$ & $\theta=0.4$ & $\alpha=0.5$ & $\lambda_1=0.3$ & $\lambda_0=0.25$ & $\rho_0=0.2$ & $\rho_1=0.35$ & $\theta=0.4$ & $\alpha=0.5$ & $\lambda_1=0.3$ & $\lambda_0=0.25$ \\
    \hline
    &  & & & & & & 
     \\
    50 & 16 &   -0.017 & -0.018 & 0.032 & 0.042 & -0.040 & 0.004 & 0.135 & 0.098 & 0.217 & 0.281 & 0.182 & 0.228  \\
     & 25&   -0.008 & -0.013 & 0.031 & 0.050 & -0.018 & -0.026  & 0.108 & 0.076 & 0.169 & 0.239 & 0.148 & 0.194  \\
     & 49 &  -0.002 & -0.007 & 0.017 & 0.028 & -0.007 & -0.018 & 0.084 & 0.052 & 0.107 & 0.171 & 0.105 & 0.156  \\
     &  & & & & & & 
     \\
    
    100 & 16 &  -0.008 & -0.005 & 0.013 & 0.053 & -0.021 & -0.021 & 0.094 & 0.070 & 0.137 & 0.240 & 0.128 & 0.175 \\
     & 25 & -0.005 & -0.004 & 0.012 & 0.042 & -0.012 & -0.022 & 0.079 & 0.054 & 0.102 & 0.197 & 0.104 & 0.148   \\
       & 49 & -0.002 & -0.002 & 0.008 & 0.023 & -0.007 & -0.013 & 0.058 & 0.035 & 0.070 & 0.128 & 0.073 & 0.116   \\
        &  & & & & & & 
     \\
    150 & 16 & -0.002 & -0.005 & 0.010 & 0.041 & -0.017 & -0.014  & 0.076 & 0.059 & 0.104 & 0.204 & 0.106 & 0.149   \\
      & 25 &  0.001 & -0.001 & 0.008 & 0.035 & -0.015 & -0.015 & 0.061 & 0.043 & 0.080 & 0.156 & 0.087 & 0.124 \\
       & 49 &  0.000 & 0.000 & 0.002 & 0.025 & -0.009 & -0.014 & 0.048 & 0.030 & 0.055 & 0.119 & 0.064 & 0.101  \\
        &  & & & & & & 
     \\
\hline
  \end{tabular}
  
\end{table}
\end{landscape}

\section{Financial network volatility: evidence from NYSE, DAX and CAC 40}\label{sec:application1}

We present an empirical application of the spatiotemporal E-GARCH model to financial network time series, specifically focusing on the stock markets of New York, Germany and France. \textcolor{blue}{The analysis uses daily price data for 30 U.S. stocks (NYSE and NASDAQ), 36 constituents of the German DAX index, and 37 constituents of the French CAC\,40 index.} This selection was guided by the availability of price data for these assets on Yahoo Finance. \textcolor{blue}{The observations span the period from 1st October 2019 to 31st October 2020, a period marked by significant shocks due to the COVID-19 crisis, hence suitable in volatility clustering and asymmetries. Each asset includes 731 daily observations across the three markets.} We consider $\xvec{Y}_t = (Y_t(s_1), \cdots, Y_t(s_n))$ where $Y_t(s_i)$ represent the log return at time $t$ of an asset $s_i$. We replace zero values with random numbers from a normal mean $0$ and variance $0.0001$, because the spatiotemporal E-GARCH model is a logarithmic volatility model and, therefore, cannot have zero observations. Replacing zero observations with small (random) numbers is frequently done in practical applications of log-GARCH models. \textcolor{blue}{Zero returns, meaning the closing price was identical to the previous day’s close, occurred in 
$0.898\%, 0.898\%, 2.675\%$ and $0.314\%$ of the observations for the CAC 40, DAX, and NYSE, respectively.}


Stock returns typically exhibit weak temporal autoregressive correlation and weak to moderate cross-sectional correlation, with the latter varying across stocks depending on their sectors. These dependencies are often removed by first fitting an autoregressive mean model and then modelling the residuals separately. In our framework, we apply a spatial dynamic panel data (SDPD) model without exogenous regressors as mean model, as defined by \cite{lee2012qml}, which is specified as
\begin{eqnarray}
\xvec{Y}_t = \rho\xvec{W}_1 \xvec{Y}_t + \gamma\xvec{Y}_{t-1} + \lambda \xvec{W}_2 \xvec{Y}_{t-1} + \xvec{u}_t , 
\end{eqnarray}
where $\rho\xvec{W}_1 \xvec{Y}_t$ captures the contemporaneous dependence across the network, $\gamma\xvec{Y}_{t-1}$ captures the temporal autoregressive effects of the own past of each stock, and $\lambda \xvec{W}_2 \xvec{Y}_{t-1}$ is the spatiotemporal spillover or diffusion effect. Eventually, the model residuals are denoted by $\xvec{u}_t$, which will be modelled as a spatiotemporal E-GARCH model in the second step. If volatility is modelled directly on raw returns without first accounting for the mean, the model may misattribute changes in the conditional mean as changes in volatility. By applying the SDPD model to filter out the predictable structure in the mean, the resulting residuals should be whitened, meaning they exhibit minimal predictable structure in terms of their conditional mean.

Unlike spatial settings, where the locations/coordinates are directly observable, the financial network structure is inherently unknown. \cite{eckel2011measuring} propose using the geographic proximity of asset headquarters, finding that distances beyond 50 miles become less relevant. Similarly, \cite{asgharian2013spatial} defines weights based on the distances between capital cities, which has been applied in studies on the US, UK, and Japanese markets. However, stocks can be traded globally with (almost) no temporal delays. Thus, the dependence often goes beyond geographical distances. \cite{fulle2022spatial} define the distance between stocks as a scaled Euclidean distance based on selected balance sheet positions. The most relevant positions are identified by optimising the goodness-of-fit of spatial GARCH models. A local optimum is found for seven key financial metrics, including free cash flow, operating expenses, and total stockholder equity. Instead of fundamental values, other approaches utilise a similarity function between the observed stock market returns. For instance, \cite{mattera2024network} considers three different distance metrics to define the financial network. More specifically, they utilise the Euclidean distance and linear correlations between the time series as well as Piccolo distances, which are defined as the Euclidean distance between the estimated coefficients of univariate log-ARCH models fitted to each time series.

\begin{figure}
\centering
  \includegraphics[width=0.32\textwidth]{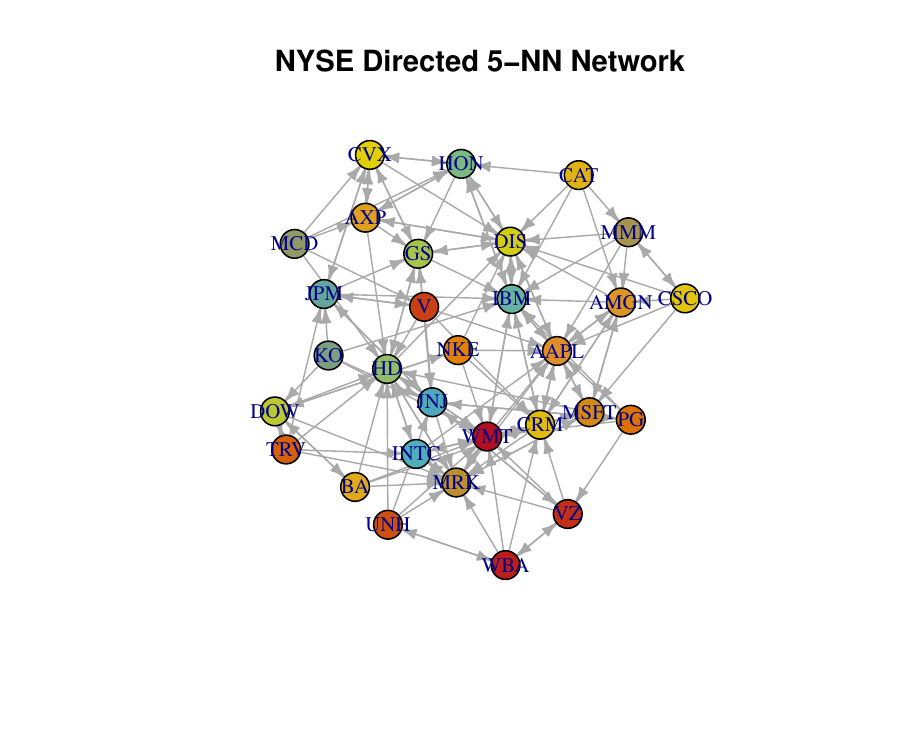}
  \includegraphics[width=0.32\textwidth]{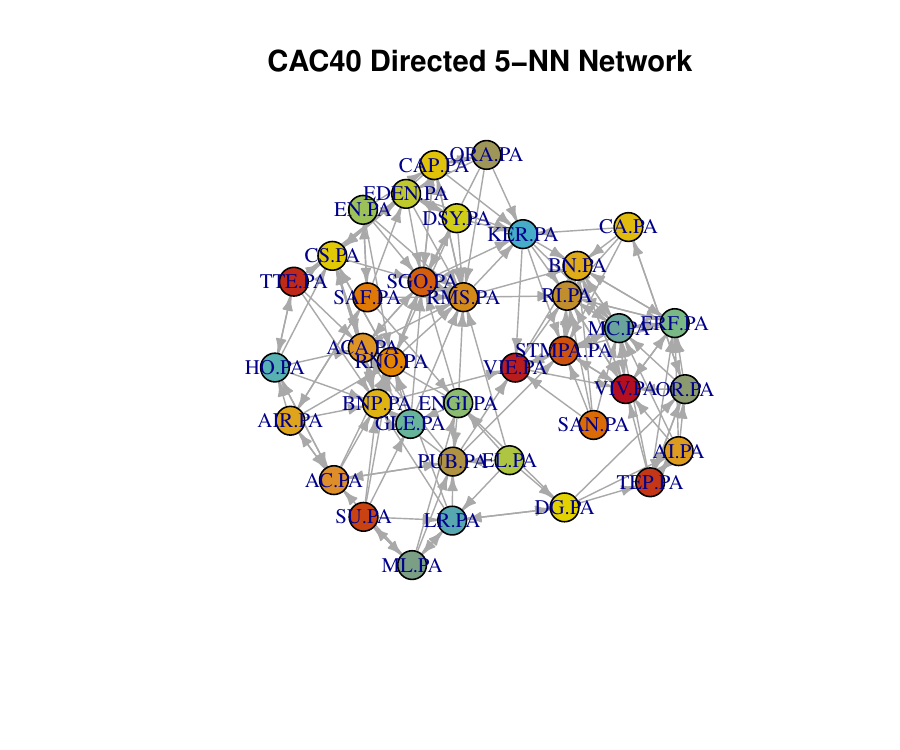}
  \includegraphics[width=0.32\textwidth]{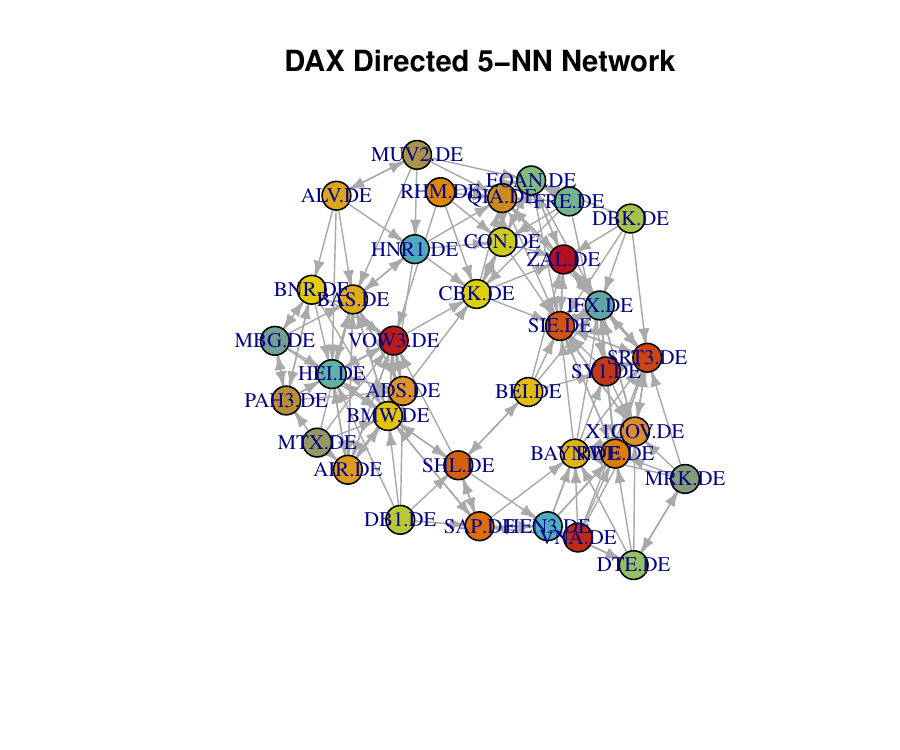}
\caption{Network constructed using the Piccolo-distance-based weight matrix for the  NYSE, CAC-40 and DAX-30 stocks (from left to right). The edges represent the 5-NN weights.}\label{fig:networks}
\end{figure}

For our analysis, we rely on these three types of network definitions considered in \cite{mattera2024network}. The weights are assigned using a $k$-nearest neighbours approach, where each node is connected to its  $k$ closest neighbours, and the weights are defined as
\begin{eqnarray}
    w_{ij} = \left \{
\begin{array}{c l}
    \frac{1}{k}  & \text{ if } i \text{ is among the } k \text{ nearest neighbours of } j \\
    0   &  \text{otherwise}
\end{array}
\right.
\end{eqnarray}
The resulting financial networks are shown in Figure \ref{fig:networks} for all three stock market examples.

Additionally, we choose two different parameters for the leverage effect, $\theta_0$ describing the leverage the effects of the contemporaneous positive and negative changes from other stocks, and $\theta_1$ for the temporal leverage effect (i.e., positive and negative changes of the same stocks in the previous period).

Optimization was performed using the Sequential Quadratic Programming (SQP) method via the `solnp' function implemented in the R-package \texttt{Rsolnp} \citep{ghalanos2012package}. The initial values were selected by generating 100 random parameter sets and choosing the one that maximised the likelihood function. The estimated parameters for each configuration, along with their corresponding standard errors, are summarised in Table \ref{tab:summary}.

The estimation results show that while the contemporaneous effect coefficient $\rho_0$ is not significant, the leverage effect  $\theta_0$ and the contemporaneous GARCH effect $\lambda_0$ are significant. This indicates that volatility exhibits direct spatial spillovers, but these effects are asymmetric, with negative shocks propagating more strongly than positive ones. This aligns with the well-documented asymmetric response of volatility to shocks, where bad news amplifies volatility spillovers across space.

Moreover, we can observe that the temporal E-GARCH effect $\rho_1$, the temporal leverage effect $\theta_1$, and the temporal GARCH effect $\lambda_1$ are all statistically significant. The moderate magnitude of $\rho_1$ suggests that past shocks contribute to volatility persistence on a moderate level. The asymmetric leverage effect $\theta_1$ is in the expected range, reinforcing the well-documented pattern that negative shocks tend to increase volatility more than positive shocks over time. The estimated GARCH effects  $\lambda_1$ are generally comparable in magnitude to the contemporaneous GARCH effects $\lambda_0$, often exceeding them but not consistently. This suggests that both spatial and temporal components play a crucial role in volatility propagation, with temporal dependence slightly dominating but not always outweighing contemporaneous spillovers. In other words, a stock’s own past changes in volatility and shocks play a more dominant role in determining its current volatility than contemporaneous effects from other stocks, though the latter remain a non-negligible factor.

Comparing the different network definitions, the distance-based and correlation-based weight matrices yield the best results in terms of AIC and BIC. Both approaches generally produce similar AIC and BIC values, whereas the Piccolo-based weights result in substantially higher AIC and BIC values, indicating a poorer model fit. \textcolor{blue}{Figures \ref{fig:appendix-nyse}, \ref{fig:appendix-cac40}, and \ref{fig:appendix-dax} in the Appendix show the daily residuals from the previously discussed SDPD model, together with their squared values, which are commonly used as a proxy for daily volatility, for the NYSE, CAC40, and DAX markets. Across all graphs, pronounced volatility clustering is observed, with clusters occurring almost simultaneously across assets and across the three markets. The corresponding estimated volatilities obtained from the spatiotemporal E-GARCH model are also reported, exhibiting high- and low-volatility periods that broadly coincide with those identified in the squared residuals. This concordance underscores the model's capacity to capture temporal variation in market volatility. For the NYSE and CAC40, the correlation-based spatial weight matrix was selected, as it delivered the best fit according to both the AIC and BIC criteria, while for the DAX, the distance-based matrix was employed.
}

\begin{table}
\centering
\begin{tabular}{ccccc}
\hline
\textbf{Network definition} & \textbf{Parameter} & \textbf{CAC40} & \textbf{DAX} & \textbf{NYSE} \\ \hline
\multirow{9}{*}{Distance-based} 
& $\rho_0$    & $\underset{(1.3631)}{0.000108}$  & $\underset{(1.456)}{0.00043}$ & $\underset{(1.584)}{0.00013 }$ \\ \cline{2-5} 
& $\rho_1$    & $\underset{(5.19 \times 10^{-3})}{0.3485  }$  & $\underset{(5.780\times 10^{-5})}{0.317 }$ & $\underset{(7.753 \times 10^{-2})}{ 0.4731 }$ \\ \cline{2-5} 
& $\theta_0$  & $\underset{(6.47 \times 10^{-3} )}{-0.2496  }$ & $\underset{( 4.370 \times 10^{-5})}{ -0.5203}$ & $\underset{ (1.547 \times 10^{-5}) }{ -0.3414}$ \\ \cline{2-5} 
& $\theta_1$  & $\underset{( 3.79 \times 10^{-2} )}{-0.1349 }$ & $\underset{( 1.24\times 10^{-3})}{ -0.1042}$ & $\underset{(8.844\times 10^{-1})}{-0.0530 }$ \\ \cline{2-5} 
& $\alpha$    & $\underset{(1.84 \times 10^{-2} )}{-0.2472 }$ & $\underset{(1.76\times 10^{-5} )}{ -0.759}$ & $\underset{(1.023\times 10^{-1} )}{-1.319}$ \\ \cline{2-5} 
& $\lambda_1$ & $\underset{( 2.39 \times 10^{-6})}{0.3991 }$  & $\underset{(2.605\times 10^{-7} )}{ 0.845}$ & $\underset{(8.155\times 10^{-7})}{ 0.5010}$ \\ \cline{2-5} 
& $\lambda_0$ & $\underset{(3.58 \times 10^{-7})}{0.5651}$  & $\underset{( 6.279 \times 10^{-7})}{ 0.059}$ & $\underset{(1.587 \times 10^{-5})}{ 0.3363 }$ \\ \cline{2-5}
& {AIC}    & \textbf{-154706.5}  &   -145380.7 & -125501.5   \\ \cline{2-5}
& {BIC}    & \textbf{-154649.1 } & -145323.4 & -125445.5  \\ \hline

\multirow{9}{*}{Correlation-based} 
& $\rho_0$    & $\underset{( 1.731)}{ 0.103}$  & $\underset{(1.7484 )}{ 0.0408}$ & $\underset{ (1.919) }{0.00084 }$ \\ \cline{2-5} 
& $\rho_1$    & $\underset{(1.46\times 10^{-2} )}{ 0.361}$  & $\underset{(2.69 \times 10^{-3} )}{0.4676 }$ & $\underset{(1.3833)}{ 0.3980}$ \\ \cline{2-5} 
& $\theta_0$  & $\underset{(2.254\times 10^{-5} )}{ 0.349}$  & $\underset{( 2.906 \times 10^{-3})}{ -0.0461}$ & $\underset{(3.847)}{-0.2790 }$ \\ \cline{2-5} 
& $\theta_1$  & $\underset{(1.31 \times 10^{-2} )}{0.0561 }$ & $\underset{(7.57 \times 10^{-3})}{-0.0264}$ & $\underset{( 2.4360)}{ 0.0249}$ \\ \cline{2-5} 
& $\alpha$    & $\underset{(4.707 \times 10^{-2} )}{-1.383 }$ & $\underset{( 2.66\times 10^{-3})}{-1.2859}$ & $\underset{(3.0586 )}{-0.5388 }$ \\ \cline{2-5} 
& $\lambda_1$ & $\underset{( 4.70 \times 10^{-2})}{0.656 }$  & $\underset{( 7.79\times 10^{-4})}{0.7444 }$ & $\underset{(1.935)}{ 0.62260}$ \\ \cline{2-5} 
& $\lambda_0$ & $\underset{(1.92 \times 10^{-6})}{ 0.178}$  & $\underset{( 1.81 \times 10^{-5})}{ 0.0921 }$ & $\underset{( 0.2760 )}{0.3088 }$ \\ \cline{2-5}
& {AIC}    &  -154498.4 &  \textbf{-145542.7} & \textbf{-126606.5} \\ \cline{2-5}
& {BIC}    &  -154487.1 & \textbf{-145531.6} & \textbf{-126596.7} \\ \hline

\multirow{9}{*}{Piccolo distance}   
& $\rho_0$    & $\underset{(1.772  )}{0.105 }$  & $\underset{( 1.043)}{ 0.00023 }$ & $\underset{(1.099)}{ 0.0016}$ \\ \cline{2-5} 
& $\rho_1$    & $\underset{(2.21\times 10^{-3} )}{ 0.297}$  & $\underset{(8.31\times 10^{-3} )}{ 0.2889}$ & $\underset{(5.030 \times 0^{-5} )}{0.3616 }$ \\ \cline{2-5} 
& $\theta_0$  & $\underset{(9.75\times 10^{-2} )}{-0.566 }$  & $\underset{(3.028 \times 10^{-5} )}{0.0275 }$ & $\underset{(1.598 \times 10^{-4} )}{0.0462 }$ \\ \cline{2-5} 
& $\theta_1$  & $\underset{(  1.388)}{ -0.026}$ & $\underset{( 8.716\times 10^{-3})}{ -0.029}$ & $\underset{(5.051 \times 10^{-3} )}{-0.0264 }$ \\ \cline{2-5} 
& $\alpha$    & $\underset{( 6.95\times 10^{-3} )}{ -0.504 }$ & $\underset{( 8.06 \times 10^{-4})}{-0.7073}$ & $\underset{( 5.0474 \times 10^{-5})}{ 0.5986 }$ \\ \cline{2-5} 
& $\lambda_1$ & $\underset{( 1.14 \times 10^{-4})}{  0.622 }$  & $\underset{( 1.53\times 10^{-3})}{ 0.909}$ & $\underset{(5.4102 \times 10^{-6})}{0.7649 }$ \\ \cline{2-5} 
& $\lambda_0$ & $\underset{( 2.48\times 10^{-3} )}{ 0.315 }$  & $\underset{( 4.47\times 10^{-7} )}{0.0017 }$ & $\underset{(5.512 \times 10^{-7} )}{0.1607 }$ \\ \cline{2-5}
& {AIC}    & -149013  & -139422.4 & -121379.7 \\ \cline{2-5}
& {BIC}    & -148955.6 & -139365.2 & -121323.7 \\ \hline
\end{tabular}
\caption{Estimates (Standard errors) and AIC/BIC for different weight matrices. Results from the STAR-SpEGARCH}
\label{tab:summary}
\end{table}

We perform diagnostic checks to assess the adequacy of the spatiotemporal E-GARCH model. The model’s performance is evaluated using autocorrelation diagnostics on both residuals and squared residuals across time and space. The Ljung-Box test, applied to each asset, indicates that only about 5\% of the assets exhibit significant autocorrelation  ($\alpha = 5\%$) in both residuals and squared residuals (maximimum 8.33\%), suggesting that the model effectively captures temporal dependencies. All p-values are reported in Figures \ref{fig:diagnostic1} to \ref{fig:diagnostic6} in the Appendix. Furthermore, Moran’s I test, conducted across all time points, shows that about 5\% of the time points (ranging from 0\% to 10.82\%) had significant correlation ($\alpha = 5\%$) in the residuals and about 9\% of the time points (ranging from 8.36\% to 11.5\%) display significant spatial autocorrelation for the squared residuals  ($\alpha = 5\%$), indicating that while some spatial dependence remains in the squared residuals, the model successfully accounts for volatility clustering in both dimensions. Moreover, we do not observe any regularities in these results across space and/or time; that is, there are no periods or node clusters showing significant correlations.

\section{Conclusion and directions for future research}\label{sec:conclusion}

This paper introduces a spatiotemporal exponential GARCH model, extending traditional volatility modelling frameworks to incorporate spatial dependencies and asymmetric spillovers. Beyond the model specification, we provide a theoretical analysis of its stochastic properties, establishing conditions for strict and weak stationarity and deriving explicit expressions for the first and second moments. These results offer fundamental insights into the behaviour of volatility in a spatiotemporal setting.

To estimate the model parameters, we propose a quasi-maximum likelihood estimator, leveraging numerical inversion techniques to approximate the unobserved innovations. The finite-sample properties of the estimator are evaluated through Monte Carlo simulations, demonstrating that the root mean squared error decreases as the number of spatial locations (nodes) or time points increases, providing empirical evidence of consistency.

We apply the spatiotemporal E-GARCH model to empirical data from the French stock market, the German stock market and the NYSE, analysing the spatiotemporal dynamics of financial volatility. The results confirm the presence of significant contemporaneous and temporal volatility spillovers, with leverage effects amplifying negative shocks more than positive ones. These findings align with established financial theory, emphasizing the asymmetric propagation of risk across assets and sectors.

Additionally, we compare different network structures for defining the spatial weight matrix. Our results indicate that distance-based and correlation-based networks provide the best fit in terms of AIC and BIC, while the Piccolo-based weighting scheme results in significantly higher information criteria values. This highlights the importance of selecting an appropriate spatial structure when modelling volatility spillovers.

For future research, several extensions could be explored. Beyond the proposed QML estimation, an alternative approach is to obtain parameter estimates via a weighted least squares procedure, minimising the weighted sum of squares
\begin{eqnarray*}
   \left( \ln \xvec{Y}_t^2 - (\xmat{I} - \lambda_0 \xmat{W}_2)^{-1} \lambda_1\ln \xvec{Y}_{t-1}^2 - E(\xvec{\nu}_t)\right)^\prime_{t=1,\cdots,T}  \left( Cov(\xvec{\nu}_t, \xvec{\nu}_s))_{t,s=1,\cdots,T} \right)^{-1} \\ \left( \ln \xvec{Y}_t^2 - (\xmat{I} - \lambda_0 \xmat{W}_2)^{-1} \lambda_1\ln \xvec{Y}_{t-1}^2 - E(\xvec{\nu}_t) \right)_{t=1,\cdots,T}  
\end{eqnarray*}
where \( ( Cov(\xvec{\nu}_t, \xvec{\nu}s))_{t,s=1,\cdots, T} \) is a block tridiagonal matrix, allowing for efficient inversion techniques \citep[see, e.g.,][]{meurant1992review}. More precisely, 
\[ ( Cov(\xvec{\nu}_t, \xvec{\nu}_s))_{t,s=1,\cdots,T} = \xmat{I}_T \otimes Cov(\xvec{\nu}_1) + \xmat{J}_T \otimes Cov(\xvec{\nu}_1, \xvec{\nu}_2) + \xmat{J}_T^\prime \otimes Cov(\xvec{\nu}_2, \xvec{\nu}_1) \]
where $\xmat{J}_T =(a_{ij})_{i,j=1,\cdots,T}$ with $a_{i,i+1} = 1$, else the elements are zero. The covariance matrix depends on the model parameters and, therefore, could get arbitrarily small if the sum of squares is directly minimised. Thus, two-step procedures (feasible least squares, \citealt{bai2021feasible}) or regularised estimators should be considered.

Additionally, based on Theorem \ref{theorem2}, we can estimate the parameters using Gaussian quasi-maximum likelihood (QML) estimation applied to the logarithmic squared observations. The corresponding log-likelihood function is given by
\begin{eqnarray*}
l(\rho_0, \rho_1, \alpha, \theta, \lambda_1, \lambda_2 | \xvec{\varepsilon}_0 , \xvec{y}_0 ) &=& -\frac{NT}{2}\ln(2\pi) - \frac{1}{2} \ln{|( Cov(\xvec{\nu}_t, \xvec{\nu}_s))_{t,s=1,\cdots,T}| }\\
&-& \frac{1}{2} \xvec{Z}_{T}^\prime  \left( Cov(\xvec{\nu}_t, \xvec{\nu}_s))_{t,s=1,\cdots,T} \right)^{-1}\xvec{Z}_{T}
\end{eqnarray*}
where \( \xvec{Z}_{T} \) represents the deviation of the log-squared process from its conditional expectation. In deriving this model, we assume that the errors follow a Gaussian distribution. However, it is well known that QML estimators generally perform well even when the error process is not Gaussian, particularly for large samples, due to their asymptotic efficiency. The transformed model simplifies the handling of nonlinearity while maintaining the structure of the residuals, which remain block tridiagonal, allowing for efficient inversion techniques. Future research could explore the robustness of this approach under alternative distributions and the potential for bias correction in small samples.

Another direction for future research could be continuous-space GARCH models, which would allow for predicting the volatility at unknown locations, similar to kriging in geostatistics. A potential point for offset could be exponential continuous-time GARCH models, as proposed by \cite{haug2007exponential}.

While our empirical analysis focuses on stock market data, future research could conduct a large-scale comparative study to examine which network structures and spatiotemporal volatility interactions are most relevant across different financial markets, including equities, commodities, and cryptocurrencies. Such an analysis could provide deeper insights into the role of spatial and temporal spillovers in diverse asset classes. Additionally, incorporating time-dynamic weight matrices—allowing for evolving financial connections based on market conditions—could enhance empirical modelling and improve the accuracy of volatility spillover detection. Investigating these extensions would contribute to a more comprehensive understanding of spatiotemporal dependencies in financial risk propagation.

\section*{Acknowledgement}
We gratefully acknowledge funding by the Deutsche Forschungsgemeinschaft (DFG, German Research Foundation) project number 501539976.

\newpage

\bibliography{references}

\newpage

\begin{appendix}

\section*{Proofs}

\begin{proof}[Theorem \ref{th:existence}]
Let $\xvec{F}_t = \ln(\xvec{h}_t)$. Then $\{ \xvec{F}_t \}$ follows a VAR(1) process. If $\phi(\lambda_1 (\xmat{I} - \lambda_0 \xmat{W}_2)^{-1}) < 1$ then the process (\ref{defF}) has a unique weakly stationary solution given by 
\[ \xvec{F}_t = \sum_{v=0}^\infty \lambda_1^v (\xmat{I} - \lambda_0 \xmat{W}_2)^{-v} \xvec{\Delta}_{t-v} . \] 
Note that
\[ E(\xvec{F}_t ) = \sum_{v=0}^\infty \lambda_1^v (\xmat{I} - \lambda_0 \xmat{W}_2)^{-v-1} \xvec{\alpha}_1 \]
and
\[ Cov(\xvec{F}_{t+h}, \xvec{F}_t) = \sum_{v,j=0}^\infty \lambda_1^{v+j} (\xmat{I} - \lambda_0 \xmat{W}_2)^{-v} Cov(\xvec{\Delta}_{t+h-v}, \xvec{\Delta}_{t-j}) (\xmat{I} - \lambda_0 \xmat{W}_2^\prime)^{-j} . \]
Because $Cov(\xvec{\Delta}_{t+h-v}, \xvec{\Delta}_{t-j}) = \xmat{0}$ if $|h-v+j| > 1$ we get that for $h \ge 0$
\begin{eqnarray*} 
Cov(\xvec{F}_{t+h}, \xvec{F}_t) & = & \sum_{v=h}^\infty \lambda_1^{2v-h} (\xmat{I} - \lambda_0 \xmat{W}_2)^{-v} Cov(\xvec{\Delta}_0, \xvec{\Delta}_0) (\xmat{I} - \lambda_0 \xmat{W}_2^\prime)^{h-v}\\ 
& & +  \sum_{v=\max\{0,h-1\}}^\infty \lambda_1^{2v-h+1} (\xmat{I} - \lambda_0 \xmat{W}_2)^{-v} Cov(\xvec{\Delta}_1, \xvec{\Delta}_0) (\xmat{I} - \lambda_0 \xmat{W}_2^\prime)^{h-v-1}  \\
& & +  \sum_{v=h+1}^\infty \lambda_1^{2v-h-1} (\xmat{I} - \lambda_0 \xmat{W}_2)^{-v} Cov(\xvec{\Delta}_0, \xvec{\Delta}_1) (\xmat{I} - \lambda_0 \xmat{W}_2^\prime)^{h-v+1} .
\end{eqnarray*}

The solution $\{ \xvec{F}_t \}$ is also strictly stationary and thus (\ref{eq:initial}) has a unique strictly stationary solution. 

Now
\begin{eqnarray*}
\xvec{F}_t & = & \sum_{v=0}^\infty \lambda_1^v (\xmat{I} - \lambda_0 \xmat{W}_2)^{-v-1} \xvec{\alpha}_1 + \rho_0 (\xmat{I} - \lambda_0 \xmat{W}_2)^{-1} \xmat{W}_1 g(\xvec{\varepsilon}_t)\\
& & + \sum_{v=1}^\infty \lambda_1^{v-1} (\xmat{I} - \lambda_0 \xmat{W}_2)^{-v} \left( \rho_0 \lambda_1 (\xmat{I} - \lambda_0 \xmat{W}_2)^{-1} \xmat{W}_1 + \rho_1 \xmat{I} \right) g(\xvec{\varepsilon}_{t-v}) .
\end{eqnarray*}
Since $\xvec{\varepsilon}_t$ is multivariate normally distributed the moment generating function of $\varepsilon_t(\xvec{s}_i)$ exists for all $t$ and $i$ and thus $E(exp(\xvec{F}_t))$ exists and is bounded. Thus all moments of $\xvec{Y}_t$ exist. Using that

\begin{eqnarray*}
Y_t(\xvec{s}_i)  & = & exp(\frac{1}{2} \sum_{v=0}^\infty \lambda_1^v \xvec{e}_i^\prime (\xmat{I} - \lambda_0 \xmat{W}_2)^{-v-1} \xvec{\alpha}_1) \\
& & \times  \; \varepsilon_t(\xvec{s}_i) \; exp\left( \frac{1}{2} \rho_0 \xvec{e}_i^\prime (\xmat{I} - \lambda_0 \xmat{W}_2)^{-1} \xmat{W}_1 g(\xvec{\varepsilon}_t) \right)\\
& & \times  \; exp\left(\frac{1}{2} \xvec{e}_i^\prime \sum_{v=1}^\infty \lambda_1^{v-1} (\xmat{I} - \lambda_0 \xmat{W}_2)^{-v} \left( \rho_0 \lambda_1 (\xmat{I} - \lambda_0 \xmat{W}_2)^{-1} \xmat{W}_1 + \rho_1 \xmat{I} \right) g(\xvec{\varepsilon}_{t-v})\right) \\
& & = I \times II \times III .
\end{eqnarray*}
Further, 
\[ I = exp\left( \frac{1}{2} \xvec{e}_i^\prime \left((1-\lambda_1) \xmat{I} - \lambda_0 \xmat{W}_2\right)^{-1} \xvec{\alpha}_1\right) . \]
We get that

\begin{eqnarray*}
E( Y_t(\xvec{s}_i) ) 
& = & I \times  \; E\left( \varepsilon_t(\xvec{s}_i) \; exp\left( \frac{1}{2} \rho_0 \xvec{e}_i^\prime (\xmat{I} - \lambda_0 \xmat{W}_2)^{-1} \xmat{W}_1 g(\xvec{\varepsilon}_t) \right) \right)\\
& \qquad \times & \prod_{v=1}^\infty  E\left(exp\left(\frac{1}{2} \xvec{e}_i^\prime \lambda_1^{v-1} (\xmat{I} - \lambda_0 \xmat{W}_2)^{-v} \left( \rho_0 \lambda_1 (\xmat{I} - \lambda_0 \xmat{W}_2)^{-1} \xmat{W}_1 + \rho_1 \xmat{I} \right) g(\xvec{\varepsilon}_{t-v})\right) \right).\\
\end{eqnarray*}

Further,
\begin{eqnarray*}
E( Y_t(\xvec{s}_i)^2 ) 
& = & I^2 \times  \; E\left( \varepsilon_t(\xvec{s}_i)^2 \; exp\left( \rho_0 \xvec{e}_i^\prime (\xmat{I} - \lambda_0 \xmat{W}_2)^{-1} \xmat{W}_1 g(\xvec{\varepsilon}_t) \right) \right)\\
& \qquad \times & \prod_{v=1}^\infty  E\left(exp\left( \xvec{e}_i^\prime \lambda_1^{v-1} (\xmat{I} - \lambda_0 \xmat{W}_2)^{-v} \left( \rho_0 \lambda_1 (\xmat{I} - \lambda_0 \xmat{W}_2)^{-1} \xmat{W}_1 + \rho_1 \xmat{I} \right) g(\xvec{\varepsilon}_{t-v})\right) \right) . 
\end{eqnarray*}
Since $\{ \xvec{\varepsilon}_t \}$ is an independent and identically distributed random sequence the distributions of $ g(\xvec{\varepsilon}_t)$ and $(\varepsilon_t(\xvec{s}_i), g(\xvec{\varepsilon_t}))$   do not depend on $t$ and we can write
\begin{eqnarray*}
E( Y_t(\xvec{s}_i)^2 ) 
& = & I^2 \times  \; E\left( \varepsilon_1(\xvec{s}_i)^2 \; exp\left( \rho_0 \xvec{e}_i^\prime (\xmat{I} - \lambda_0 \xmat{W}_2)^{-1} \xmat{W}_1 g(\xvec{\varepsilon}_1) \right) \right)\\
& \qquad \times & \prod_{v=1}^\infty  E\left(exp\left( \xvec{e}_i^\prime \lambda_1^{v-1} (\xmat{I} - \lambda_0 \xmat{W}_2)^{-v} \left( \rho_0 \lambda_1 (\xmat{I} - \lambda_0 \xmat{W}_2)^{-1} \xmat{W}_1 + \rho_1 \xmat{I} \right) g(\xvec{\varepsilon}_{1})\right) \right) . 
\end{eqnarray*}

Next we consider the autocovariances. We use the notation $I_i$, $II_i$, and $III_i$ as above but referring to the index $i$ and $I_j$, $II_j$, $III_j$ for the index $j$. Then, 
\begin{eqnarray*}
E( Y_t(\xvec{s}_i) Y_t(\xvec{s}_j) ) & = & I_i I_j E( II_i II_j) E(III_i III_j)\\
& = & exp\left( \frac{1}{2} (\xvec{e}_i + \xvec{e}_j)^\prime \left((1-\lambda_1) \xmat{I} - \lambda_0 \xmat{W}_2\right)^{-1} \xvec{\alpha}_1\right)\\
& & \times \;  E\left( \varepsilon_1(\xvec{s}_i) \varepsilon_1(\xvec{s}_j) \; exp\left( \frac{1}{2} \; \rho_0 (\xvec{e}_i + \xvec{e}_j)^\prime (\xmat{I} - \lambda_0 \xmat{W}_2)^{-1} \xmat{W}_1 g(\xvec{\varepsilon}_1) \right) \right)\\
& \qquad \times & \prod_{v=1}^\infty  E\left(exp\left( \frac{1}{2} \; (\xvec{e}_i + \xvec{e}_j)^\prime \lambda_1^{v-1} (\xmat{I} - \lambda_0 \xmat{W}_2)^{-v} \left( \rho_0 \lambda_1 (\xmat{I} - \lambda_0 \xmat{W}_2)^{-1} \xmat{W}_1 + \rho_1 \xmat{I} \right) g(\xvec{\varepsilon}_{1-v})\right) \right) .
\end{eqnarray*}
As above we conclude that this quantity does not depend on $t$. \\

Now let $\xi = 0$. We use that if $X \sim \Phi$ that 
\[ E(exp(aX)) = exp(a^2/2), E(X exp(aX)) = a \, exp(a^2/2), E(X^2 exp(aX)) = (1+a) exp(a^2/2) .\] 
Further, if $X_1,...,X_n$ are independent and standard normally distributed then 
\begin{eqnarray*}
E(exp(\sum_{i=1}^n a_i X_i)) & = & \prod_{i=1}^n exp(a_i^2/2) = exp(\sum_{i=1}^n a_i^2/2) , \\
E(X_1 exp(\sum_{i=1}^n a_i X_i)) & = & a_1 exp(\sum_{i=1}^n a_i^2/2) ,\\
E(X_1^2 exp(\sum_{i=1}^n a_i X_i)) & = & (1+a_1^2) exp(\sum_{i=1}^n a_i^2/2) ,\\
E( X_1 X_2 exp(\sum_{i=1}^n a_i X_i)) & = & a_1 a_2 exp(\sum_{i=1}^n a_i^2/2) . 
\end{eqnarray*}
Consequently,
\[ E\left(exp\left(\frac{1}{2} \xvec{e}_i^\prime \lambda_1^{v-1} (\xmat{I} - \lambda_0 \xmat{W}_2)^{-v} \left( \rho_0 \lambda_1 (\xmat{I} - \lambda_0 \xmat{W}_2)^{-1} \xmat{W}_1 + \rho_1 \xmat{I} \right) g(\xvec{\varepsilon}_{t-v})\right) \right) \] 
\[ = exp(\frac{1}{8} \Theta^2 \lambda_1^{2v-2} \xvec{e}_i^\prime  (\xmat{I} - \lambda_0 \xmat{W}_2)^{-v} \left( \rho_0 \lambda_1 (\xmat{I} - \lambda_0 \xmat{W}_2)^{-1} \xmat{W}_1 + \rho_1 \xmat{I} \right) \left( \rho_0 \lambda_1 (\xmat{I} - \lambda_0 \xmat{W}_2)^{-1} \xmat{W}_1 + \rho_1 \xmat{I} \right)^\prime (\xmat{I} - \lambda_0 \xmat{W}_2^\prime)^{-v} \xvec{e}_i )  \]
and 
\[ E\left( \varepsilon_t(\xvec{s}_i) \; exp\left( \frac{1}{2} \rho_0 \xvec{e}_i^\prime (\xmat{I} - \lambda_0 \xmat{W}_2)^{-1} \xmat{W}_1 g(\xvec{\varepsilon}_t) \right) \right) \]
\[ = \frac{1}{2} \rho_0 \Theta \xvec{e}_i^\prime (\xmat{I} - \lambda_0 \xmat{W}_2)^{-1} \xmat{W}_1 \xvec{e}_i \times \, exp(\frac{1}{8} \rho_0^2 \Theta^2 \xvec{e}_i^\prime (\xmat{I} - \lambda_0 \xmat{W}_2)^{-1} \xmat{W}_1 \xmat{W}_1^\prime (\xmat{I} - \lambda_0 \xmat{W}_2^\prime)^{-1} \xvec{e}_i ) . \]
Moreover, 
\[ E\left( \varepsilon_t(\xvec{s}_i)^2 \; exp\left( \rho_0 \xvec{e}_i^\prime (\xmat{I} - \lambda_0 \xmat{W}_2)^{-1} \xmat{W}_1 g(\xvec{\varepsilon}_t) \right) \right) \]
\[ = \left( 1+  \rho_0^2 \Theta^2 (\xvec{e}_i^\prime (\xmat{I} - \lambda_0 \xmat{W}_2)^{-1} \xmat{W}_1 \xvec{e}_i)^2 \right) \times \, exp( \frac{1}{2} \rho_0^2 \Theta^2 \xvec{e}_i^\prime (\xmat{I} - \lambda_0 \xmat{W}_2)^{-1} \xmat{W}_1 \xmat{W}_1^\prime (\xmat{I} - \lambda_0 \xmat{W}_2^\prime)^{-1} \xvec{e}_i ) , \]
\[ E\left(exp\left( \xvec{e}_i^\prime \lambda_1^{v-1} (\xmat{I} - \lambda_0 \xmat{W}_2)^{-v} \left( \rho_0 \lambda_1 (\xmat{I} - \lambda_0 \xmat{W}_2)^{-1} \xmat{W}_1 + \rho_1 \xmat{I} \right) g(\xvec{\varepsilon}_{t-v})\right) \right) \] 
\[ = exp\left(\frac{1}{2} \Theta^2 \lambda_1^{2v-2} \xvec{e}_i^\prime  (\xmat{I} - \lambda_0 \xmat{W}_2)^{-v} \left( \rho_0 \lambda_1 (\xmat{I} - \lambda_0 \xmat{W}_2)^{-1} \xmat{W}_1 + \rho_1 \xmat{I} \right) \left( \rho_0 \lambda_1 (\xmat{I} - \lambda_0 \xmat{W}_2)^{-1} \xmat{W}_1 + \rho_1 \xmat{I} \right)^\prime (\xmat{I} - \lambda_0 \xmat{W}_2^\prime)^{-v} \xvec{e}_i \right)  . \]
Furthermore,
\begin{eqnarray*}
E\left( \varepsilon_1(\xvec{s}_i) \varepsilon_1(\xvec{s}_j) \; exp\left( \frac{1}{2} \; \rho_0 (\xvec{e}_i + \xvec{e}_j)^\prime (\xmat{I} - \lambda_0 \xmat{W}_2)^{-1} \xmat{W}_1 g(\xvec{\varepsilon}_1) \right) \right) & = & \frac{1}{4} \Theta^2 \rho_0^2 (\xvec{e}_i + \xvec{e}_j)^\prime (\xmat{I} - \lambda_0 \xmat{W}_2)^{-1} \xmat{W}_1  \xvec{e}_i \times \\
& & \hspace*{-10cm} \times (\xvec{e}_i + \xvec{e}_j)^\prime (\xmat{I} - \lambda_0 \xmat{W}_2)^{-1} \xmat{W}_1  \xvec{e}_j  \times exp\left( \frac{1}{8} \rho_0^2 \Theta^2   (\xvec{e}_i + \xvec{e}_j)^\prime (\xmat{I} - \lambda_0 \xmat{W}_2)^{-1} \xmat{W}_1 \xmat{W}_1^\prime (\xmat{I} - \lambda_0 \xmat{W}_2^\prime)^{-1} (\xvec{e}_i + \xvec{e}_j) \right) , 
\end{eqnarray*}
\begin{eqnarray*}
E\left(exp\left( \frac{1}{2} (\xvec{e}_i + \xvec{e}_j)^\prime \lambda_1^{v-1} (\xmat{I} - \lambda_0 \xmat{W}_2)^{-v} \left( \rho_0 \lambda_1 (\xmat{I} - \lambda_0 \xmat{W}_2)^{-1} \xmat{W}_1 + \rho_1 \xmat{I} \right) g(\xvec{\varepsilon}_{1})\right) \right) & = &  \\
& & \hspace*{-13cm} exp\left( \frac{1}{4} \Theta^2 \lambda_1^{2v-2} (\xvec{e}_i + \xvec{e}_j)^\prime (\xmat{I} - \lambda_0 \xmat{W}_2)^{-v} \left( \rho_0 \lambda_1 (\xmat{I} - \lambda_0 \xmat{W}_2)^{-1} \xmat{W}_1 + \rho_1 \xmat{I} \right) \times \right. \\
& & \hspace*{-13cm} \left. \times \left( \rho_0 \lambda_1 (\xmat{I} - \lambda_0 \xmat{W}_2)^{-1} \xmat{W}_1 + \rho_1 \xmat{I} \right)^\prime (\xmat{I} - \lambda_0 \xmat{W}_2^\prime)^{-v}
(\xvec{e}_i + \xvec{e}_j) \right) .
\end{eqnarray*}

\end{proof}

\begin{proof}[Theorem \ref{theorem2}]
Under the assumption that $\xvec{\varepsilon}_t$ has zero mean and constant variance, we have $\xvec{\nu}_t= \ln \xvec{Y}_t^2 - (\xmat{I} - \lambda_0 \xmat{W}_2)^{-1} \lambda_1\ln \xvec{Y}_{t-1}^2 $ has a finite variance.
This is proved using that
 \begin{equation}
	\ln \xvec{Y}_t^2 = \ln \xvec{h}_t + \ln \xvec{\varepsilon}_t^2 \, ,
\end{equation}
and we get, 
 \begin{equation}
	\ln \xvec{Y}_t^2 = \mathbf{S} \alpha_1 + \ln \xvec{\varepsilon}_t^2 + \mathbf{S}\rho_0 \xmat{W}_1 g(\xvec{\varepsilon}_t) + \mathbf{S} \rho_1 g(\xvec{\varepsilon}_{t-1}) + \mathbf{S}\lambda_1\ln \xvec{Y}_{t-1}^2 - \mathbf{S}\lambda_1\ln \xvec{\varepsilon}_{t-1}^2   \, .
\end{equation}
Where $\mathbf{S}=\mathbf{S}_n(\lambda_0) = (\xmat{I} - \lambda_0 \xmat{W}_2)^{-1}$.
Now, let
 \begin{eqnarray*}
	\xvec{\nu}_t & = &  \ln \xvec{Y}_t^2 - (\xmat{I} - \lambda_0 \xmat{W}_2)^{-1} \lambda_1\ln \xvec{Y}_{t-1}^2 \\
	& = &  (\xmat{I} - \lambda_0 \xmat{W}_2)^{-1} (\alpha\xvec{1} + \rho_0 \xmat{W}_1 g(\xvec{\varepsilon}_t) + \rho_1 g(\xvec{\varepsilon}_{t-1}) + \lambda_1 \ln \xvec{\varepsilon}_{t-1}^2) + \ln \xvec{\varepsilon}_t^2 \, .
\end{eqnarray*}
$\xvec{\nu}_t$ is written as a function as a finite combination of $\xvec{\varepsilon}_{t-1}$ and $\xvec{\varepsilon}_{t}$.   Because $\xvec{\varepsilon}$ has zero mean and constant variance, in this sense we conclude that $\xvec{\nu}_t$ has a constant variance as well.

a) Now 
 \begin{eqnarray*}
	E(\xvec{\nu}_t) 
 & = &  (\xmat{I} - \lambda_0 \xmat{W}_2)^{-1}\xvec{\alpha}_1  + (-ln(2) - \gamma ) (\xmat{I} - \lambda_1 (\xmat{I} - \lambda_0 \xmat{W}_2)^{-1}) \xvec{1} 
\end{eqnarray*}
since $E(ln \; \epsilon_t(\xvec{s}_i)^2) = - ln \; 2 - \gamma \approx - 1.27036...$ (e.g., Pav (2015)). If further $\xmat{W}_2$ is row-standardized then 
\begin{eqnarray*}
	E(\xvec{\nu}_t) 
 & = &  (\xmat{I} - \lambda_0 \xmat{W}_2)^{-1}\xvec{\alpha}_1  + (-ln(2) - \gamma ) 
 (1-\frac{\lambda_1}{1-\lambda_0}) \xvec{1} .  
\end{eqnarray*}

b) 
\[
    Cov(\xvec{\nu}_t) = Cov( \rho_0 (\xmat{I} - \lambda_0 \xmat{W}_2)^{-1} \xmat{W}_1 g(\xvec{\varepsilon}_t) + \ln \xvec{\varepsilon}_t^2) + Cov((\xmat{I} - \lambda_0 \xmat{W}_2)^{-1} 
    (\rho_1 g(\xvec{\varepsilon}_{t-1})  - \lambda_1 \ln \xvec{\varepsilon}_{t-1}^2) ) = I + II.
\]
Now, 
\begin{eqnarray*}
   I & = & \rho_0^2  (\xmat{I} - \lambda_0 \xmat{W}_2)^{-1} \xmat{W}_1
   E(g(\xvec{\varepsilon}_t) g(\xvec{\varepsilon}_t)^\prime ) \xmat{W}_1^\prime(\xmat{I} - \lambda_0 \xmat{W}_2^\prime)^{-1} + Cov( \ln \xvec{\varepsilon}_t^2 )\\
   & & + \rho_0 (\xmat{I} - \lambda_0 \xmat{W}_2)^{-1} \xmat{W}_1 Cov(g(\xvec{\varepsilon}_t),\ln \xvec{\varepsilon}_t^2) + \rho_0 Cov(\ln \xvec{\varepsilon}_t^2,g(\xvec{\varepsilon}_t)) \xmat{W}_1^\prime(\xmat{I} - \lambda_0 \xmat{W}_2^\prime)^{-1} \\
   & & = III + IV + V . 
   \end{eqnarray*}
As shown in (\ref{varg}) it holds that $Cov(g(\xvec{\varepsilon}_t)) = (\Theta^2 + \xi^2 (1-2/\pi)) \xmat{I}$. Further, we get with Pav (2015) and 8.366-9 of Gradshteyn et al. (1994) that $Cov(\ln \xvec{\varepsilon}_t^2) = \Psi^\prime(1/2) \, \xmat{I} = \pi^2/2 \, \xmat{I}$ where $\Psi(x)$ denotes the digamma function. Moreover, $E(\varepsilon_t(\xvec{s}_i) \ln \varepsilon_t(\xvec{s}_i)^2 ) = 0$ since $\varepsilon_t(\xvec{s}_i)$ is symmetric.  Finally, using 4.331-1 of Gradshteyn et al. (1994) we have that 
\begin{eqnarray*}
E(|\varepsilon_t(\xvec{s}_i)| \ln \varepsilon_t(\xvec{s}_i)^2 ) & = &  \frac{1}{\sqrt{2\pi}} \int_{-\infty}^\infty |x| \ln x^2 exp(-x^2/2) dx\\
& = & \sqrt{\frac{2}{\pi}} \int_{0}^\infty (\ln u + \ln 2) e^{-u} du = \sqrt{\frac{2}{\pi}} (\ln 2 - \gamma)  
\end{eqnarray*} 
and using that $E( |\varepsilon_t(\xvec{s}_i)| ) = \sqrt{2/\pi}$

\begin{eqnarray*}
    Cov(\ln \varepsilon_t(\xvec{s}_i)^2,g(\varepsilon_t(\xvec{s}_i))) & = & \xi E( (\ln \varepsilon_t(\xvec{s}_i)^2 + \ln 2 + \gamma) (|\varepsilon_t(\xvec{s}_i)| - \sqrt{\frac{2}{\pi}}))  =  2 \xi \ln 2 \sqrt{\frac{2}{\pi}}  ,
\end{eqnarray*} 
consequently $Cov(\ln \xvec{\varepsilon}_t^2,g(\xvec{\varepsilon}_t)) = 2 \xi \ln 2 \sqrt{\frac{2}{\pi}} \xmat{I}$.

Putting these results together we get
\begin{eqnarray*}
I & = & \rho_0^2 (\Theta^2 + \xi^2(1-\frac{2}{\pi})) (\xmat{I} - \lambda_0 \xmat{W}_2)^{-1} \xmat{W}_1 \xmat{W}_1^\prime(\xmat{I} - \lambda_0 \xmat{W}_2^\prime)^{-1} + \frac{\pi^2}{2} \; \xmat{I} \\
& & + \rho_0  2 \xi \ln 2 \sqrt{\frac{2}{\pi}} \left( (\xmat{I} - \lambda_0 \xmat{W}_2)^{-1} \xmat{W}_1 + \xmat{W}_1^\prime(\xmat{I} - \lambda_0 \xmat{W}_2^\prime)^{-1} \right) \\
\end{eqnarray*}
By analogy,
\begin{eqnarray*}
   II & = &  (\xmat{I} - \lambda_0 \xmat{W}_2)^{-1} \left( \rho_1^2 
   E(g(\xvec{\varepsilon}_t) g(\xvec{\varepsilon}_t)^\prime) + \lambda_1^2  Cov( \ln \xvec{\varepsilon}_t^2 )  - 2 \rho_1 \lambda_1  Cov(g(\xvec{\varepsilon}_t),\ln \xvec{\varepsilon}_t^2) \right)  (\xmat{I} - \lambda_0 \xmat{W}_2^\prime)^{-1} \\
   & = & \left(\rho_1^2 (\Theta^2 + \xi^2 (1- \frac{2}{\pi})) + \lambda_1^2 \frac{\pi^2}{2} - 4 \ln 2 \rho_1 \lambda_1 \xi \sqrt{\frac{2}{\pi}}\right) 
   (\xmat{I} - \lambda_0 \xmat{W}_2)^{-1}(\xmat{I} - \lambda_0 \xmat{W}_2^\prime)^{-1} .
\end{eqnarray*}

c) \begin{eqnarray*}
Cov(\xvec{\nu}_t, \xvec{\nu}_{t-1}) & = & Cov\left(  (\xmat{I} - \lambda_0 \xmat{W}_2)^{-1} \rho_0 \xmat{W}_1 g(\xvec{\varepsilon}_t) + 
\ln \xvec{\varepsilon}_t^2 + (\xmat{I} - \lambda_0 \xmat{W}_2)^{-1} (
\rho_1 g(\xvec{\varepsilon}_{t-1}) - \lambda_1 \ln \xvec{\varepsilon}_{t-1}^2), \right.\\
& & \left. ( (\xmat{I} - \lambda_0 \xmat{W}_2)^{-1} \rho_0 \xmat{W}_1 g(\xvec{\varepsilon}_{t-1}) + 
\ln \xvec{\varepsilon}_{t-1}^2 + (\xmat{I} - \lambda_0 \xmat{W}_2)^{-1} (
\rho_1 g(\xvec{\varepsilon}_{t-2}) - \lambda_1 \ln \xvec{\varepsilon}_{t-2}^2) \right) \\
& = & Cov( (\xmat{I} - \lambda_0 \xmat{W}_2)^{-1} (
\rho_1 g(\xvec{\varepsilon}_{t-1}) - \lambda_1 \ln \xvec{\varepsilon}_{t-1}^2),(\xmat{I} - \lambda_0 \xmat{W}_2)^{-1} \rho_0 \xmat{W}_1 g(\xvec{\varepsilon}_{t-1}) + 
\ln \xvec{\varepsilon}_{t-1}^2 )\\
& = & (\xmat{I} - \lambda_0 \xmat{W}_2)^{-1} \; Cov( 
\rho_1 g(\xvec{\varepsilon}_{t-1}) - \lambda_1 \ln \xvec{\varepsilon}_{t-1}^2,(\xmat{I} - \lambda_0 \xmat{W}_2)^{-1} \rho_0 \xmat{W}_1 g(\xvec{\varepsilon}_{t-1}) + 
\ln \xvec{\varepsilon}_{t-1}^2 )\\
& = &  \rho_0 \rho_1 (\xmat{I} - \lambda_0 \xmat{W}_2)^{-1}  Cov(g(\xvec{\varepsilon}_{t-1})) \xmat{W}_1^\prime (\xmat{I} - \lambda_0 \xmat{W}_2^\prime)^{-1} + (\xmat{I} - \lambda_0 \xmat{W}_2)^{-1} \rho_1 
Cov(g(\xvec{\varepsilon}_{t-1}), \ln \xvec{\varepsilon}_{t-1}^2) \\
& & \hspace*{-0.5cm} - \lambda_1 \rho_0 (\xmat{I} - \lambda_0 \xmat{W}_2)^{-1} Cov(\ln \varepsilon_{t-1}^2, g(\xvec{\varepsilon}_{t-1})) \xmat{W}_1^\prime (\xmat{I} - \lambda_0 \xmat{W}_2^\prime)^{-1} - \lambda_1 (\xmat{I} - \lambda_0 \xmat{W}_2)^{-1} Cov(\ln \xvec{\varepsilon}_{t-1}^2 ) \\
& = & \rho_0 \rho_1 (\Theta^2 + \xi^2 (1 - \frac{2}{\pi})) (\xmat{I} - \lambda_0 \xmat{W}_2)^{-1} \xmat{W}_1^\prime (\xmat{I} - \lambda_0 \xmat{W}_2^\prime)^{-1} + 2 \rho_1 \xi \ln 2 \sqrt{\frac{2}{\pi}} (\xmat{I} - \lambda_0 \xmat{W}_2)^{-1}\\
& & 
 - 2 \lambda_1 \rho_0 \xi \ln 2 \sqrt{\frac{2}{\pi}} (\xmat{I} - \lambda_0 \xmat{W}_2)^{-1} \xmat{W}_1^\prime (\xmat{I} - \lambda_0 \xmat{W}_2^\prime)^{-1} -  \lambda_1 \frac{\pi^2}{2} (\xmat{I} - \lambda_0 \xmat{W}_2)^{-1}\\
& = & \left( \rho_0 \rho_1 (\Theta^2 + \xi^2 (1 - \frac{2}{\pi})) - 2 \lambda_1 \rho_0 \xi \ln 2 \sqrt{\frac{2}{\pi}}\right) \; (\xmat{I} - \lambda_0 \xmat{W}_2)^{-1} \xmat{W}_1^\prime (\xmat{I} - \lambda_0 \xmat{W}_2^\prime)^{-1}\\
& & + \left(2 \rho_1 \xi \ln 2 \sqrt{\frac{2}{\pi}} - \lambda_1 \frac{\pi^2}{2}\right) 
\; (\xmat{I} - \lambda_0 \xmat{W}_2)^{-1} . 
\end{eqnarray*}
d) $\xvec{\nu}_t$ and $\xvec{\nu}_{t-s}$ are independent for $s > 1$ and thus the covariance matrix is zero.

\end{proof}

\begin{proof}[Theorem \ref{th:inversion}]

Now
\begin{equation}\label{start} \xvec{F}_t = \sum_{v=0}^{t-1} \lambda_1^v (\xmat{I} - \lambda_0 \xmat{W}_2)^{-v} \xvec{\Delta}_{t-v} + \lambda_1^t (\xmat{I} - \lambda_0 \xmat{W}_2)^{-t} \xvec{F}_0  \end{equation}
is a function of $\xvec{\varepsilon}_t$,..., $\xvec{\varepsilon}_0$, and $\xvec{F}_0$. Further, $\xvec{F}_0 = \ln( \xvec{Y}_0 \odot \xvec{\varepsilon}_0^{-1})$ provided that all elements of $\xvec{\varepsilon}_0$ are unequal to zero what happens with probability 1. 

Assume that $\xvec{Y}_0$ and $\xvec{\varepsilon}_0$ are known.
Then, $\xvec{Y}_1$ is a function of $\xvec{\varepsilon}_1$, i.e. $\xvec{Y}_1 = f_1(\xvec{\varepsilon}_1)$. Following the inverse function theorem the function $f_1$ is invertible if the determinant of the corresponding Jacobi matrix is unequal to zero. Now assuming that $\varepsilon_1(\xvec{s}_i) \neq 0$ for all $i=1,..,n$ it follows that
\[ \frac{\partial Y_1(\xvec{s}_i)}{\partial \varepsilon_1(\xvec{s}_j)} = \sqrt{h_1(\xvec{s}_i)} \left( \delta_{ij} + \frac{1}{2} \varepsilon_1(\xvec{s}_i) \, \frac{\partial \ln h_1(\xvec{s}_i)}{\partial \varepsilon_1(\xvec{s}_j)} \right) . \] 
With the notation $(\xvec{b}_i^\prime) = \rho_0 (\xmat{I} - \lambda_0 \xmat{W}_2)^{-1} \xmat{W}_1 = (b_{ij})$ we get that
\[ \frac{\partial \ln h_1(\xvec{s}_i)}{\partial \varepsilon_1(\xvec{s}_j)} = \xvec{b}_i^\prime ( \Theta \xvec{e}_j + \xi sgn(\varepsilon_1(\xvec{s}_j)) \xvec{e}_j) = b_{ij} ( \Theta + \xi  sgn(\varepsilon_1(\xvec{s}_j)) ) \]
where $\xvec{e}_j$ denotes the n-dimensional vector whose $j$th component is equal to $1$ and all ones are $0$ and sgn stands for the sign function. Consequently, the inverse exists if 
\[ \det \left( \xmat{I} + \frac{1}{2} \rho_0 (\xmat{I} - \lambda_0 \xmat{W}_2)^{-1} \xmat{W}_1 \odot ( \Theta \xvec{\varepsilon}_1 \xvec{1}^\prime + \xi \xvec{\varepsilon}_1  sgn(\xvec{\varepsilon}_1)^\prime)\right) \neq 0 . \]
If this is fulfilled then we can solve the equation with respect to $\xvec{\varepsilon}_1$.

Next we consider the case $t=2$. We have to solve \begin{equation}\label{t=2}
\xvec{Y}_2 = \xvec{\varepsilon}_2 \odot \sqrt{\xvec{h}_2} = \tilde{f}_2(\xvec{\varepsilon}_1, \xvec{\varepsilon}_2) . \end{equation}
with respect to $\xvec{\varepsilon}_1$ and $\xvec{\varepsilon}_2$. However, since $\xvec{\varepsilon}_1$ is known from the first equation we have to solve (\ref{t=2}) only with respect to $\xvec{\varepsilon}_2$, i.e. $\xvec{Y}_2 = f_2(\xvec{\varepsilon}_2)$. This is done with the same argumentation as for $t=1$ and we get that the inverse exists if
\[ det\left( \xmat{I} + \frac{1}{2} \rho_0 (\xmat{I} - \lambda_0 \xmat{W}_2)^{-1} \xmat{W}_1 \odot ( \Theta \xvec{\varepsilon}_2 \xvec{1}^\prime + \xi \xvec{\varepsilon}_2  sgn(\xvec{\varepsilon}_1)^\prime )\right) \neq 0 . \]

The procedure is continued for $t > 2$ and the proof is finished.
\end{proof}

\section{Additional graphics for diagnostic checks}

\begin{figure}
\centering
  \includegraphics[width=1\textwidth]{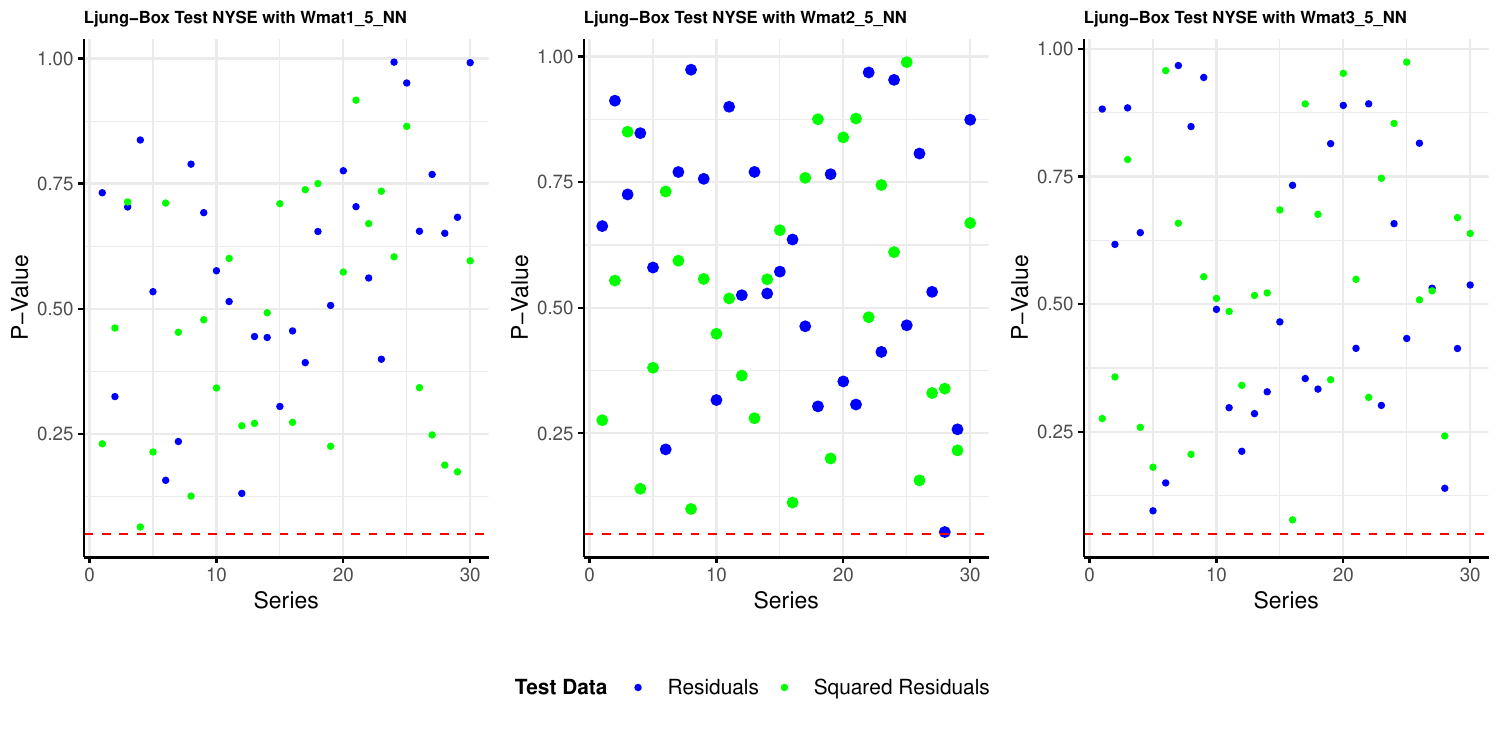}
\caption{P-values from the Ljung-Box test applied to residuals and squared residuals for the NYSE. The percentage of series with p-values above 0.05, indicating no significant autocorrelation, is as follows: Euclidean distance-based matrix (residuals: 8.33\%, squared residuals: 2.77\%), correlation-based matrix (0\%, 5.55\%), and model-based (Piccolo distance) matrix (8.33\%, 2.77\%).} \label{fig:diagnostic1}
\end{figure}

\begin{figure}
\centering
  \includegraphics[width=1\textwidth]{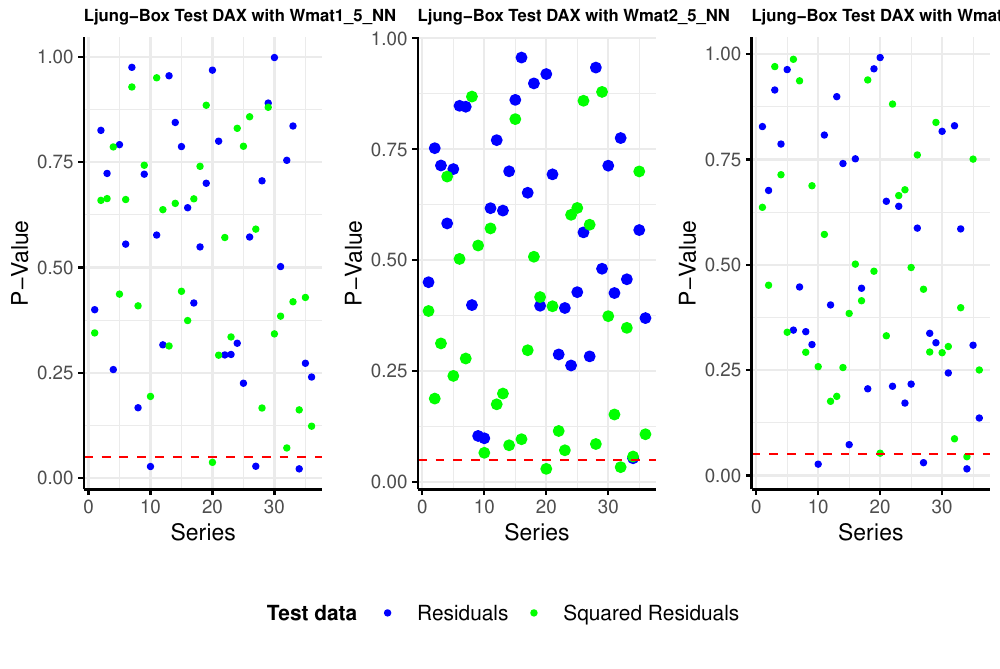}
\caption{P-values from the Ljung-Box test applied to residuals and squared residuals for the DAX-30. The percentage of series with p-values above 0.05, indicating no significant autocorrelation, is as follows: Euclidean distance-based matrix (residuals: 8.33\%, squared residuals: 5.55\%), correlation-based matrix (0\%, 2.77\%), and model-based (Piccolo distance) matrix (8.33\%, 2.77\%).} \label{fig:diagnostic2}
\end{figure}

\begin{figure}
\centering
  \includegraphics[width=1\textwidth]{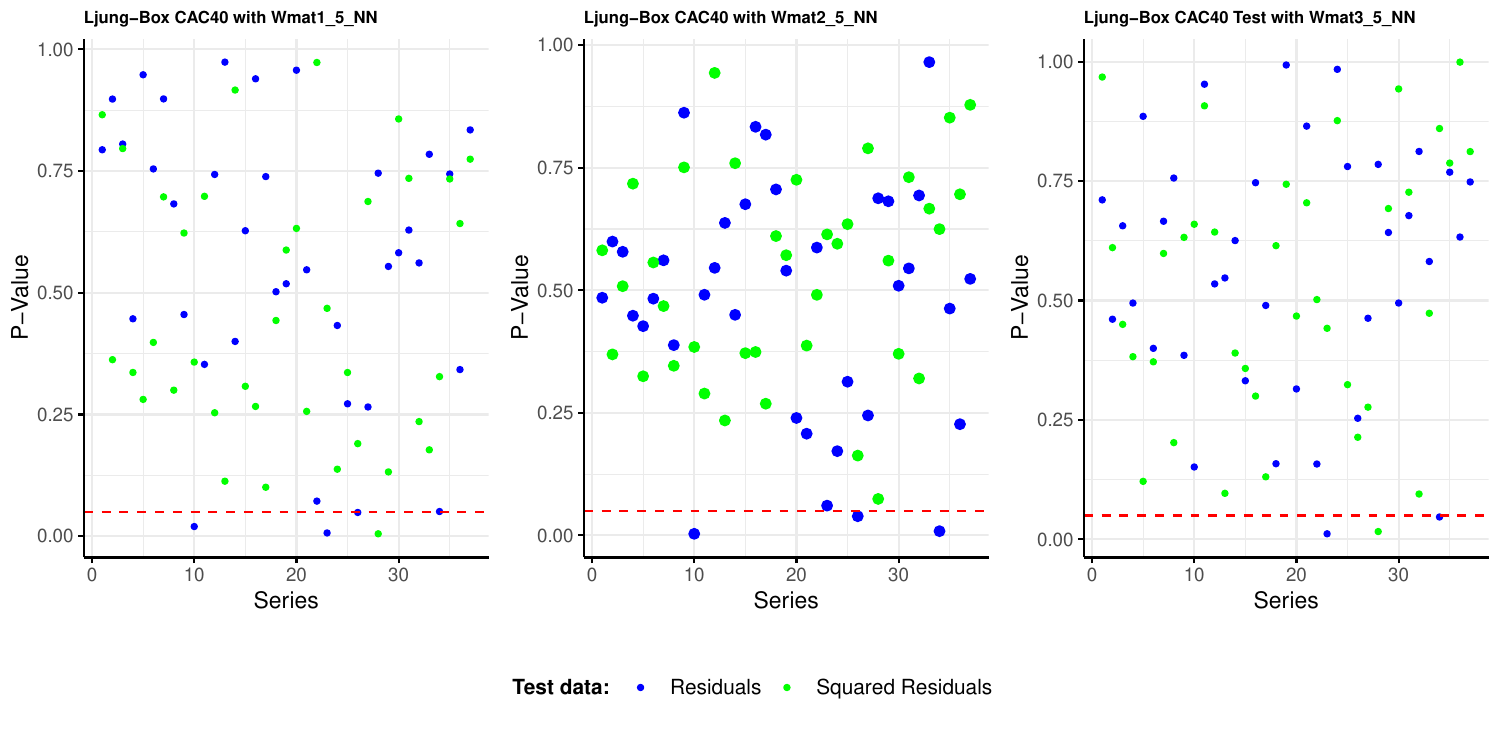}
\caption{P-values from the Ljung-Box test applied to residuals and squared residuals for the CAC-40. The percentage of series with p-values above 0.05, indicating no significant autocorrelation, is as follows: Euclidean distance-based matrix (residuals: 8.108\%, squared residuals: 2.702\%), correlation-based matrix (8.108\%, 0\%), and model-based (Piccolo distance) matrix (5.405\%, 2.702\%).}  \label{fig:diagnostic3}
\end{figure}

\begin{figure}
\centering
  \includegraphics[width=1\textwidth]{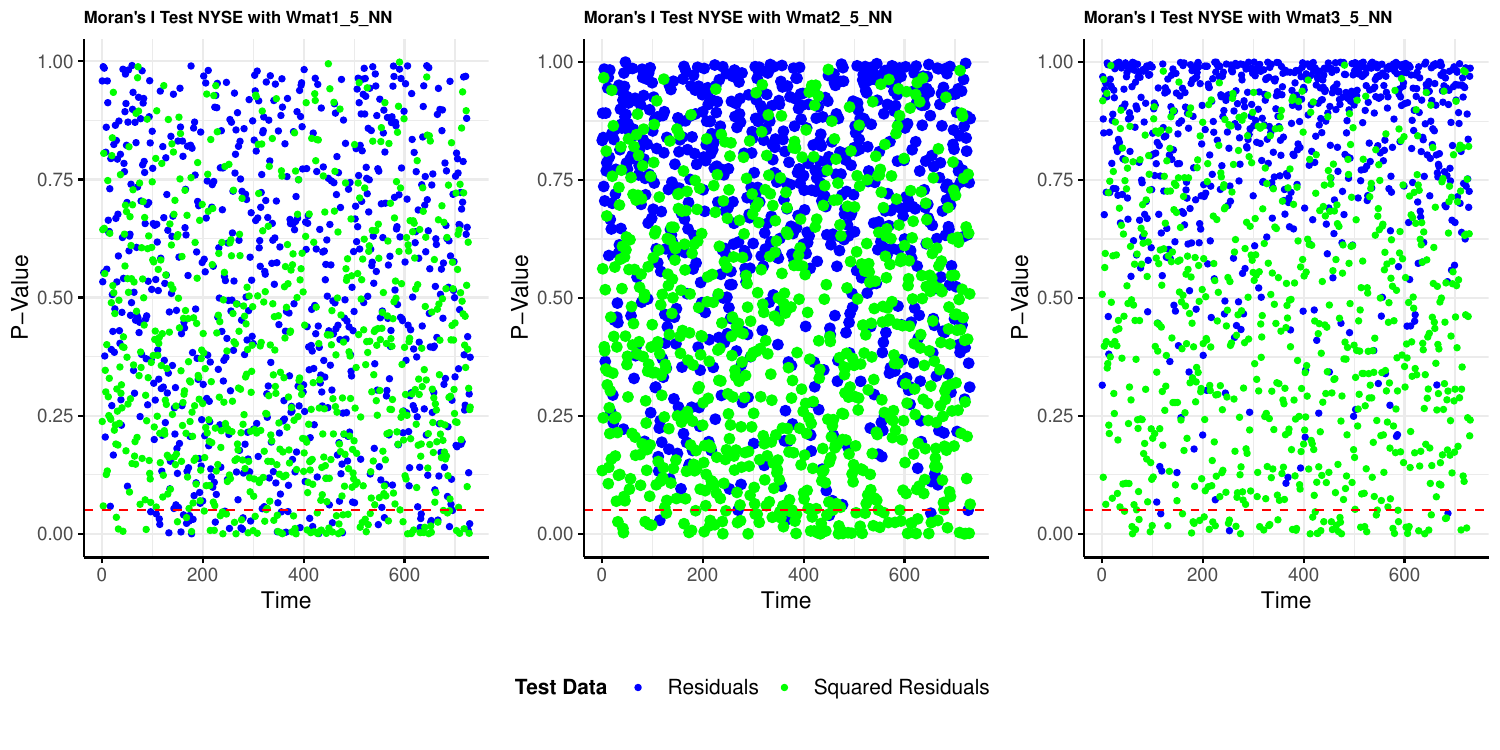}
\caption{P-values from Moran’s I test applied to residuals and squared residuals for the NYSE. The percentage of time points with p-values above 0.05, indicating no significant spatial autocorrelation, is as follows: Euclidean distance-based matrix (residuals: 5.62\%, squared residuals: 10.00\%), correlation-based matrix (0.68\%, 9.73\%), and model-based (Piccolo distance) matrix (0.41\%, 8.49\%).}  \label{fig:diagnostic4}
\end{figure}

\begin{figure}
  \includegraphics[width=1\textwidth]{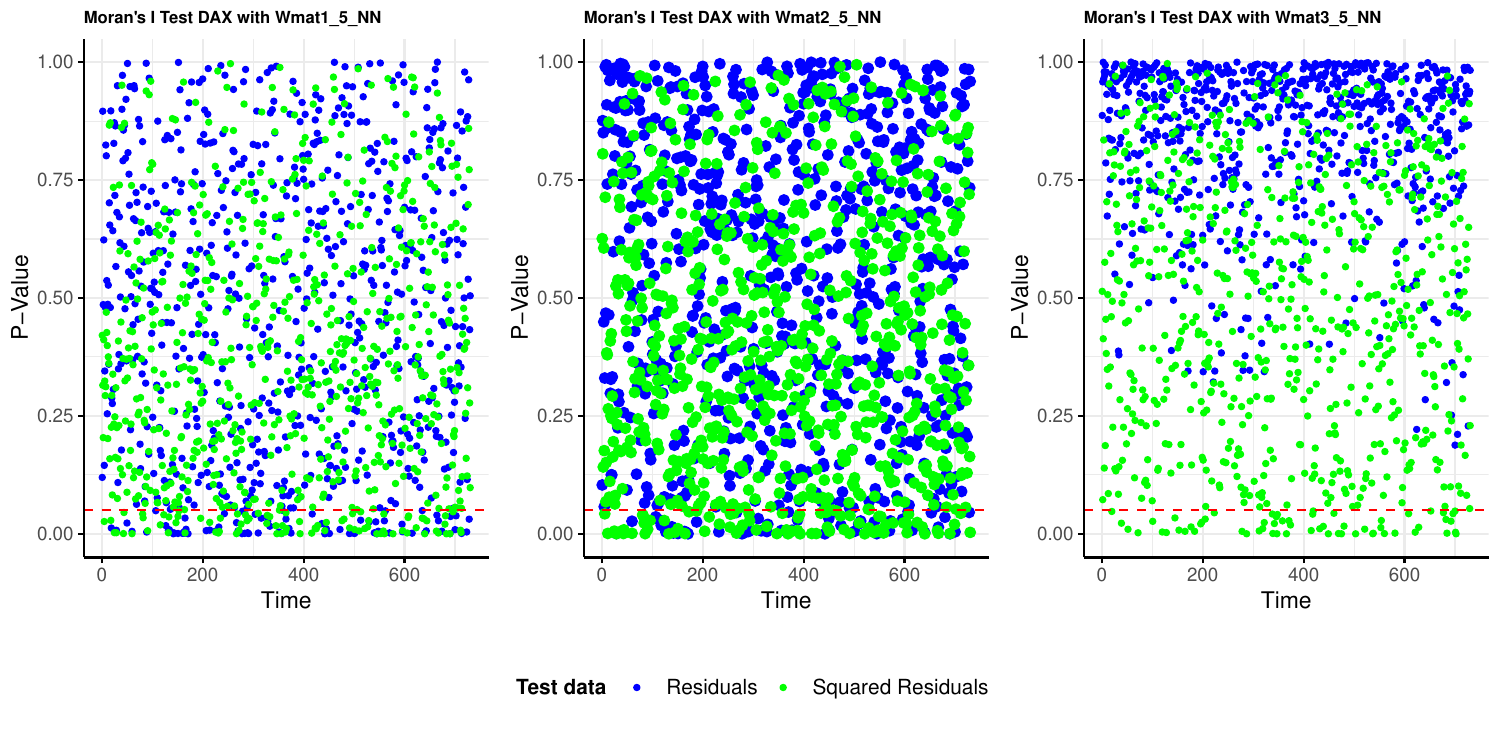}
\caption{P-values from Moran’s I test applied to residuals and squared residuals for the DAX-30. The percentage of time points with p-values above 0.05, indicating no significant spatial autocorrelation, is as follows: Euclidean distance-based matrix (residuals: 10.54\%, squared residuals: 11.50\%), correlation-based matrix (7.39\%, 11.09\%), and model-based (Piccolo distance) matrix (0\%, 8.36\%).} \label{fig:diagnostic5}
\end{figure}

\begin{figure}
\centering
  \includegraphics[width=1\textwidth]{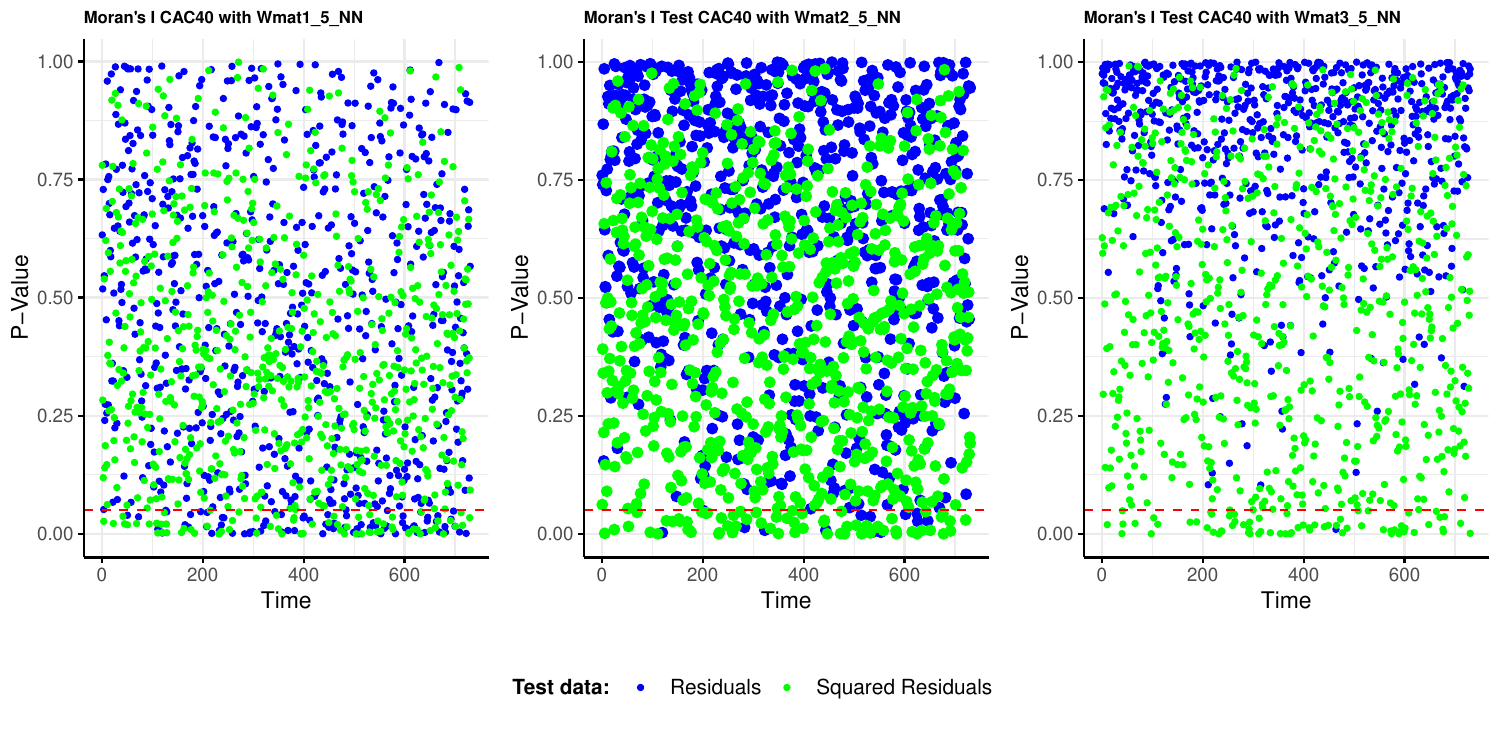}
\caption{P-values from Moran’s I test applied to residuals and squared residuals for the CAC-40. The percentage of time points with p-values above 0.05, indicating no significant spatial autocorrelation, is as follows: Euclidean distance-based matrix (residuals: 10.82\%, squared residuals: 9.58\%), correlation-based matrix (1.91\%, 8.90\%), and model-based (Piccolo distance) matrix (0.136\%, 9.31\%).}  \label{fig:diagnostic6}
\end{figure}

\begin{landscape}
\begin{figure}[p] 
  \centering
  \begin{minipage}{0.65\textwidth}
    \includegraphics[width=\linewidth]{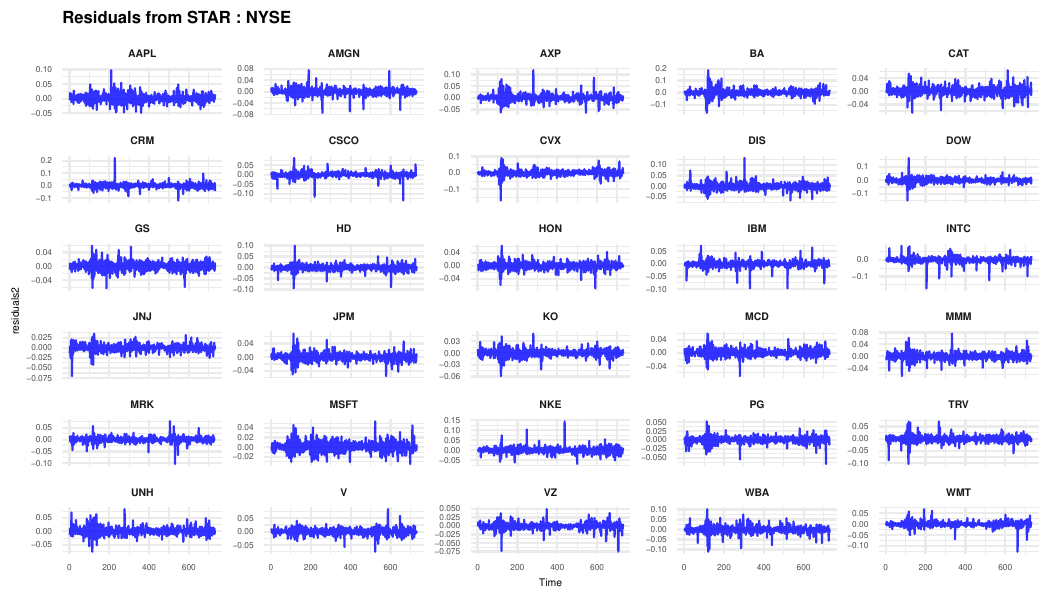}
  \end{minipage}\hspace{-0.2em}
  \begin{minipage}{0.65\textwidth}
    \includegraphics[width=\linewidth]{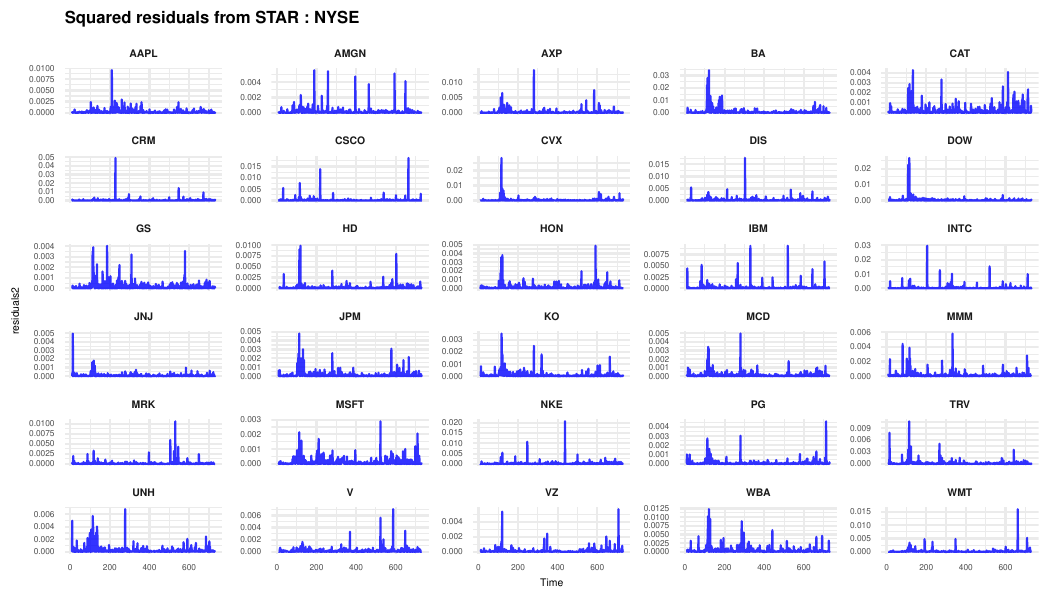}
  \end{minipage}

  \vspace{0.5em}

  \begin{minipage}{0.65\textwidth}
    \includegraphics[width=\linewidth]{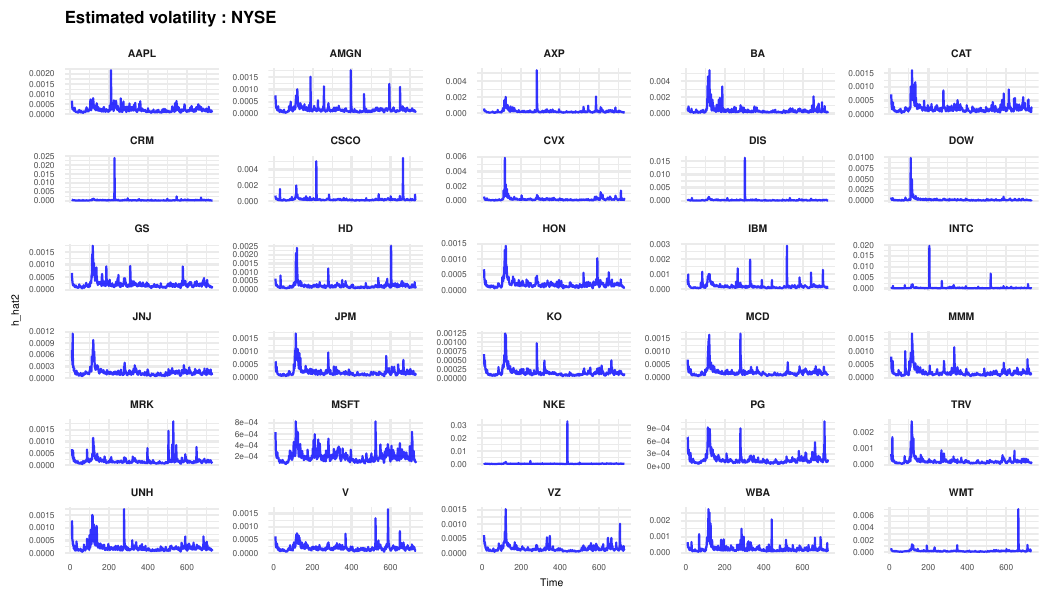}
  \end{minipage}

  \caption{STAR (Spatiotemporal Autoregressive) SDPD residuals (top left), their squared values (top right), and spatiotemporal E-GARCH volatility estimates (middle bottom) for the NYSE.}
  \label{fig:appendix-nyse}
\end{figure}  
\end{landscape}

\begin{landscape}
\begin{figure}[p]
  \centering
  \begin{minipage}{0.65\textwidth}
    \includegraphics[width=\linewidth]{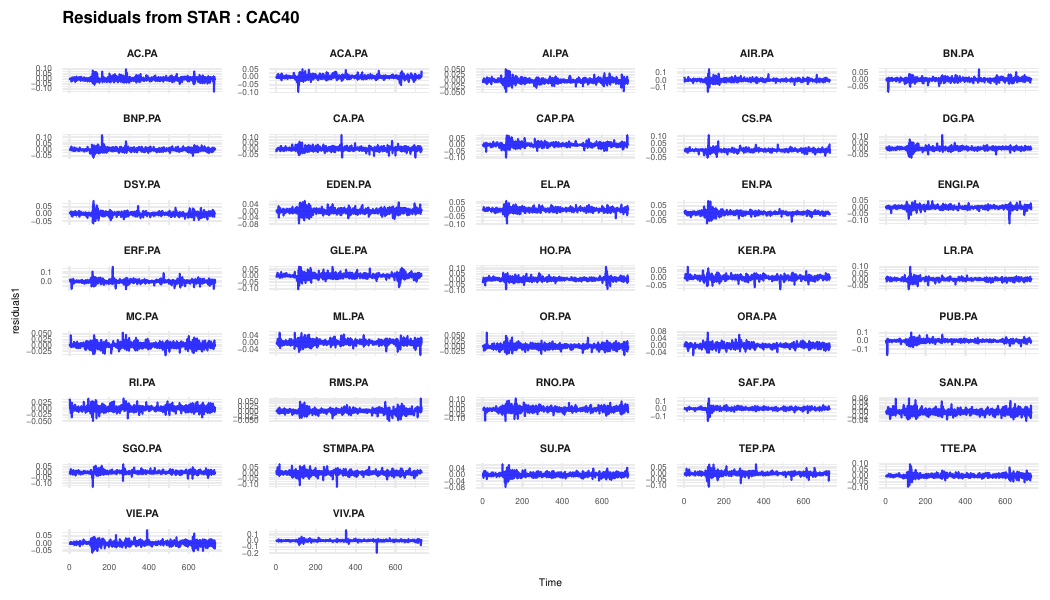}
  \end{minipage}\hspace{-0.2em}
  \begin{minipage}{0.65\textwidth}
    \includegraphics[width=\linewidth]{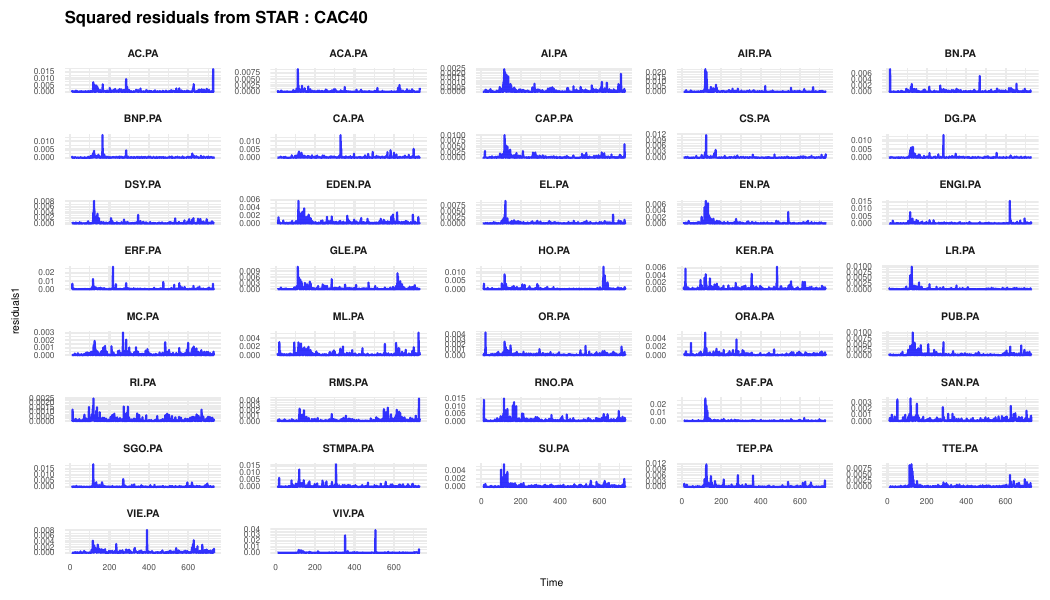}
  \end{minipage}

  \vspace{0.5em}

  \begin{minipage}{0.65\textwidth}
    \includegraphics[width=\linewidth]{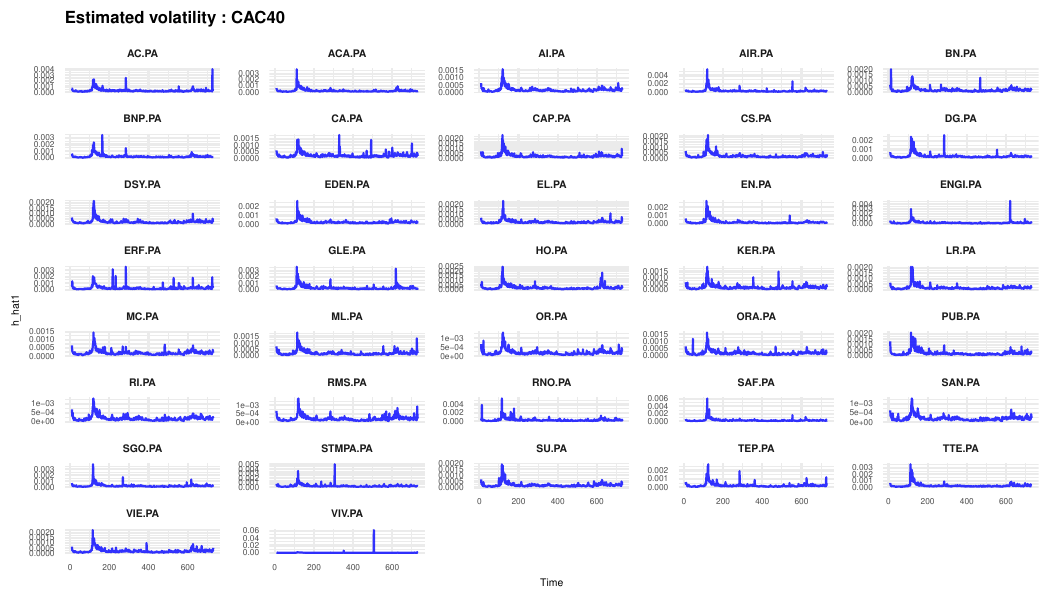}
  \end{minipage}

  \caption{STAR (Spatiotemporal Autoregressive) SDPD residuals (top left), their squared values (top right), and spatiotemporal E-GARCH volatility estimates (middle bottom) for the CAC 40.}
  \label{fig:appendix-cac40}
\end{figure}  
\end{landscape}

\begin{landscape}
\begin{figure}[p]
  \centering
  \begin{minipage}{0.65\textwidth}
    \includegraphics[width=\linewidth]{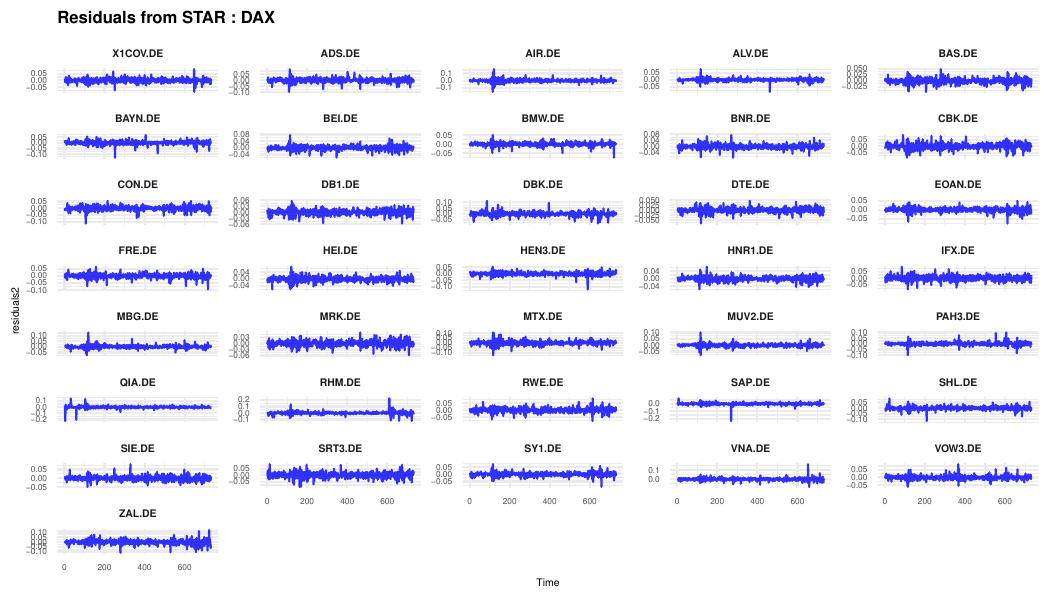}
  \end{minipage}\hspace{-0.2em}
  \begin{minipage}{0.65\textwidth}
    \includegraphics[width=\linewidth]{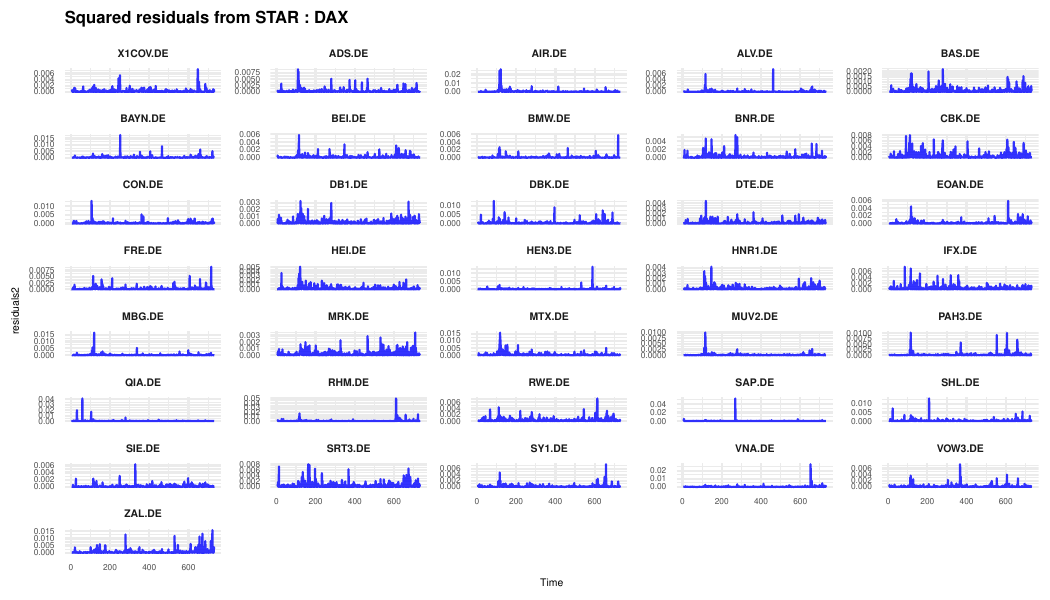}
  \end{minipage}

  \vspace{0.5em}

  \begin{minipage}{0.65\textwidth}
    \includegraphics[width=\linewidth]{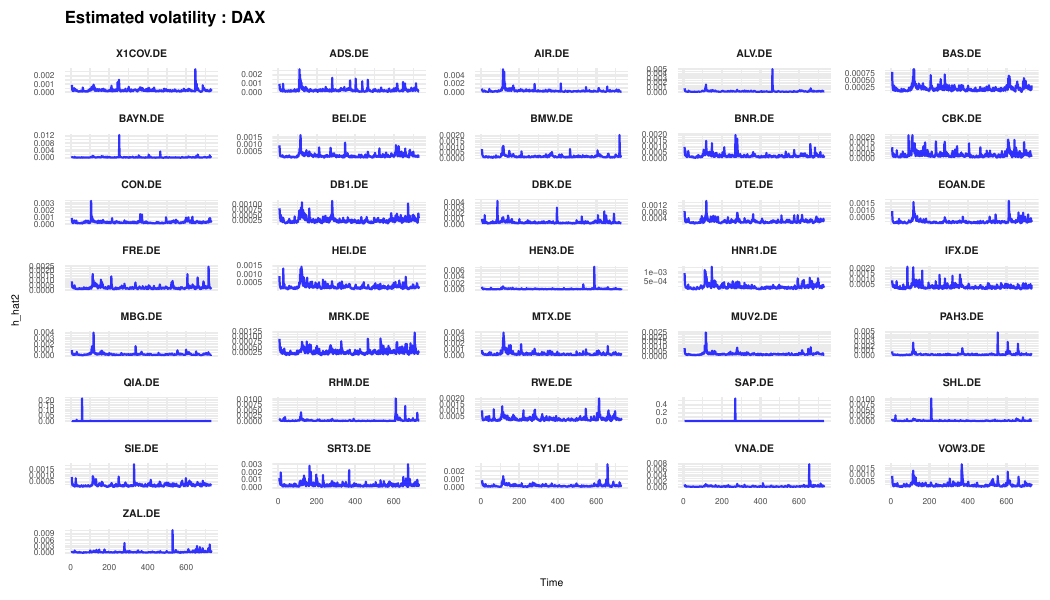}
  \end{minipage}

  \caption{STAR (Spatiotemporal Autoregressive) SDPD residuals (top left), their squared values (top right), and spatiotemporal E-GARCH volatility estimates (middle bottom) for the DAX.}
  \label{fig:appendix-dax}
\end{figure}  
\end{landscape}

\end{appendix}

\end{document}